\documentclass[11pt,a4paper]{article}

\usepackage{geometry}
\usepackage{indentfirst}

\usepackage{amsmath}
\usepackage{amssymb}
\usepackage{physics}
\usepackage{comment}

\usepackage{array}
\usepackage{booktabs}
\usepackage{longtable}
\usepackage{makecell}
\usepackage{tabularx}

\usepackage{graphicx}
\usepackage{float}
\usepackage{caption}
\usepackage{xcolor}
\usepackage{listings}
\definecolor{LightGray}{gray}{0.95}
\usepackage{csquotes}
\usepackage[
    backend=biber,
    style=numeric, 
    sorting=none]
{biblatex}
\renewbibmacro{in:}{%
    \ifentrytype{article}
    {}
    {\printtext{\bibstring{in}\intitlepunct}}}
\usepackage{hyperref}
\hypersetup{
    colorlinks=true,                               
    linktoc=all,                                    
    linkcolor=black,                                  
    citecolor=black,
    urlcolor=black,
    pdftitle={A Three-Party W-State Quantum Secret Sharing Protocol with X-Gate Encoding and Forbidden-Outcome Detection},
    pdfauthor={Timur R. Teregulov and Elena R. Loubenets},
}

\begin{document}
\title{A Three-Party W-State Quantum Secret Sharing Protocol with X-Gate Encoding and Forbidden-Outcome Detection}
\author{Timur R. Teregulov$^{1}$ and Elena R. Loubenets$^{1}$
\\ 
\\
$^{1}$Department of Applied Mathematics, MIEM, HSE University, \\
Moscow 123458, Russia\\}
\date{}
\maketitle

\begin{abstract}
We propose a new three-party quantum secret sharing (QSS) protocol based on a three-qubit W state for distributing a predetermined sequence of classical bits. Alice encodes each bit using X gates and randomly applies Hadamard gates to the transmitted qubits. Her Hadamard-gate choices and the measurement bases independently chosen by Bob and Charlie determine the type of each round—information, test or semi-information—during transmission, without designating test rounds in advance. Suitable semi-information rounds provide additional opportunities for secret transmission and improve efficiency. The protocol combines secret transmission and tests for forbidden outcomes within the same ensemble of encoded W states. In the ideal model without an attack, transmitting an $n$-bit secret requires an average of $3n$ protocol rounds, giving a qubit efficiency of $1/9$. We analyze intercept-and-resend and entangle-and-measure attacks by an internal adversary and derive their single-round and cumulative detection probabilities. For the entangle-and-measure attack, we relate the information obtainable through an ancillary system to the probability of forbidden test outcomes. The protocol is implemented in Python using Qiskit and evaluated through ideal simulation and experiments on IBM Quantum hardware. The simulation results are consistent with the theoretical predictions, while hardware measurements characterize round efficiency, reconstruction errors, forbidden-outcome probabilities and the information obtained through the specified ancilla measurement. A calibrated statistical criterion accounts for forbidden outcomes caused by hardware noise, with detection and false-alarm probabilities evaluated using a binomial model with measured rates. This analysis connects the ideal protocol predictions with a finite-sample assessment under the specified attack models and statistical assumptions.
\end{abstract}

\section{Introduction}

Secret sharing distributes a secret among several participants so that only authorized subsets can reconstruct it by pooling their shares. Shamir introduced a classical construction of this principle in 1979 \cite{Shamir1979}. Although classical secret sharing can provide information-theoretic security, protecting the distribution of shares remains a separate communication problem. Quantum cryptography provides methods for detecting eavesdropping through disturbances of transmitted quantum systems, as demonstrated by the BB84 and entanglement-based quantum key distribution protocols \cite{BennettBrassard1984,Ekert1991}. The vulnerability of widely used public-key mechanisms to quantum algorithms further motivates the study of cryptographic approaches based on quantum principles \cite{Shor1994}. These approaches extend beyond key distribution to tasks such as quantum secure direct communication and quantum secret sharing \cite{DengLongLiu2003,Hillery1999}.

Hillery, Bu{\v{z}}ek and Berthiaume introduced a three-party quantum secret sharing (QSS) protocol using Greenberger--Horne--Zeilinger (GHZ) states \cite{Hillery1999}, while Karlsson, Koashi and Imoto investigated secret sharing and secret splitting using quantum entanglement \cite{Karlsson1999}. Subsequent studies developed protocols without entanglement, efficient multiparty schemes and constructions based on local operations \cite{Guo2003,Xiao2004,Gao2005LocalOperations,ZhangLiMan2005}. Other approaches employed entanglement swapping, tripartite entangled states and higher-dimensional systems \cite{ZhangMan2005,XueYiCao2006,ZhangLiuFangWang2007}. Multiparty sharing using two specially chosen entangled states was also investigated \cite{TsaiHwang2012}. These developments established a range of QSS constructions with different resource requirements and sharing procedures.

Another direction concerns the quantum capabilities required of the participants. The semi-quantum model, originally introduced for quantum key distribution, allows some participants to perform only a restricted set of quantum operations \cite{Boyer2007}. This model was subsequently applied to secret sharing, leading to semi-quantum secret sharing (SQSS) schemes with restricted-capability participants \cite{LiChanLong2010,LiQiuMateus2013}. Within this framework, Xie et al. proposed sharing predetermined bits using GHZ-type multipartite states \cite{Xie2015Specific}, while Tian et al. developed a Bell-state protocol for the same task \cite{Tian2021Specific}. Further constructions employ cluster and Bell states \cite{Li2021Cluster}, $\chi$-type states \cite{Chen2022Chi} or a circular transmission structure combining single particles with GHZ-type states \cite{Hou2024Circular}.

Within this broader range of entangled resources, W states provide a distinct basis for constructing secret-sharing protocols. D{\"u}r, Vidal and Cirac showed that the three-qubit W and GHZ states belong to inequivalent classes of genuine tripartite entanglement and possess different entanglement properties \cite{DurVidalCirac2000}. Joo et al. investigated W-state-based quantum secure communication and partial secret sharing \cite{Joo2005WState}, while Liu et al. developed a multipartite QSS protocol using symmetric W states \cite{LiuTsaiHwang2012W}. Kaur and Kumar also investigated a W-state construction in which a controller's measurement outcome distinguishes message and control modes \cite{KaurKumar2018}. W states were subsequently employed in three-party SQSS \cite{TsaiYangLee2019W}, anonymous quantum secret sharing \cite{LiAnonymousW2024} and semi-quantum sharing of predetermined bits \cite{Xing2025W}. These studies illustrate how the same class of states can support different communication tasks and participant models.

For any such construction, security depends on the encoding, public announcements and verification procedure, rather than on the choice of entangled resource alone. An internal adversary is particularly relevant to QSS because a dishonest participant already possesses a legitimate share of the distributed system. Participant attacks have demonstrated that some protocols allow information to be obtained without introducing the errors expected by their verification procedures \cite{Qin2007HBB,Song2009ParticipantAttack}. Security vulnerabilities were also identified in the predetermined-bit schemes of Xie et al. and Tian et al., motivating corresponding improvements \cite{YinFu2016Xie,He2024}. Other studies have examined Trojan horse attacks and intercept-and-resend strategies, together with corresponding countermeasures \cite{DengTrojan2005,LinYangTsaiHwang2013,YangTsaiHwang2012}. Entangle-and-measure attacks introduce a further possibility: an adversary couples an intercepted system to an ancilla and retains the ancilla for a later measurement. Such interactions have been analyzed in QSS and related schemes \cite{QinTsoDai2019,Chou2021,Long2021,He2024}. Their effect must be assessed using the measurement statistics and information available in the particular protocol.

Experimental studies provide a complementary assessment of QSS implementations. Early demonstrations used single-qubit and multiphoton approaches, four-party entanglement and telecommunication fibers \cite{Chen2005,Schmid2005,Gaertner2007,Bogdanski2008}. Subsequent experiments explored multipartite bound entanglement, polarization-entangled photon pairs and multiparty communication systems \cite{Zhou2018,Williams2019,Liu2023,Qin2024}. QSS has also been demonstrated in a three-node superconducting quantum network \cite{Yan2026}. In practical implementations, however, noise can produce outcomes that are excluded by an ideal protocol model. Assessing such outcomes therefore requires distinguishing the ideal verification rule from a statistical comparison with the background behavior of the hardware.

These considerations motivate a protocol-specific analysis connecting secret encoding, round selection and observable disturbances. For a W-state construction, it is necessary to identify which measurement configurations permit secret recovery, which provide valid security tests and how an adversary's intervention changes their outcome probabilities. A software implementation must reproduce these choices and correlations, while a hardware analysis must account for deviations already present without an attack.

In this work, we introduce a new three-party QSS protocol based on a three-qubit W state for distributing a predetermined sequence of classical bits. Alice encodes each bit using a randomly selected permitted configuration of X gates and independently applies a Hadamard gate to each transmitted qubit with probability $1/2$. The central idea is that Alice's Hadamard-gate choices, together with the measurement bases independently chosen by Bob and Charlie, determine the type of each round: information, test or semi-information. The rounds used for testing are therefore determined during transmission rather than designated in advance and are announced by Alice after transmission. Suitable semi-information rounds are also used for secret transmission, improving efficiency. Unlike the W-state schemes of Liu et al. and Xing et al., which use separate decoy photons for security checks \cite{LiuTsaiHwang2012W,Xing2025W}, the proposed protocol tests for forbidden outcomes using the encoded W states themselves. To the best of our knowledge, this combination has not previously been proposed in W-state QSS protocols for predetermined classical secrets. The quantum communication proceeds from Alice to the receivers, who perform local measurements in the Z or X basis.

We derive the expected resource consumption and analyze intercept-and-resend and entangle-and-measure attacks performed by an internal adversary on another participant's qubit, obtaining the corresponding single-round and cumulative detection probabilities. For the entangle-and-measure attack, we relate the information that Bob can obtain about Charlie's qubit through the ancillary system to the probability of forbidden outcomes. The analysis focuses on this information-extraction mechanism, without considering joint operations involving Bob's own qubit.

The protocol is implemented in Python using the Qiskit framework. The implementation combines ideal simulation with quantum processor experiments to evaluate round efficiency, the bit error rate of the reconstructed secret, forbidden-outcome probabilities and the information obtained through the specified ancilla measurement. A calibrated statistical criterion is used to assess excess forbidden outcomes relative to the hardware background. We also compare selected resource and functional characteristics of the proposed scheme with existing QSS and SQSS protocols.

The remainder of the paper is organized as follows. Section~\ref{sec:2} presents the proposed protocol and the different types of rounds. Section~\ref{sec:3} analyzes its efficiency and examines the two internal attack models. Section~\ref{sec:4} describes the Qiskit implementation and the analysis of simulator and quantum-processor measurements. Section~\ref{sec:5} compares the proposed scheme with selected existing protocols. Finally, Section~\ref{sec:6} concludes the paper.

\section{Proposed protocol}\label{sec:2}

In this section, we propose a new quantum secret sharing protocol based on W states \cite{DurVidalCirac2000}. The main idea is that the dealer uses the X gate to encode the secret bits and randomly applies the Hadamard gate to the qubits before sending them to the participants as an additional security measure, while the participants independently choose their measurement basis.

The dealer, hereafter referred to as Alice, first prepares a three-qubit W state, since the proposed scheme is designed for two secret recipients:
\begin{equation}
    \ket{W}_{CBA}=\frac{1}{\sqrt{3}}(\ket{001}+\ket{010}+\ket{100}). \label{W_orig}
\end{equation} 
Alice retains qubit A, while qubits B and C are assigned to Bob and Charlie, respectively. To encode each secret bit, Alice applies one of the permitted combinations of X gates and then independently applies the Hadamard gate to each of the qubits assigned to Bob and Charlie with probability 1/2. Alice then sends the corresponding qubits to the participants, who independently choose between the Z and X measurement bases with equal probability, perform their measurements and subsequently announce their chosen bases over an authenticated public channel. Based on the chosen bases and the Hadamard gates she applied, Alice classifies each round into one of three types: information, test or semi-information. After transmitting the entire sequence of states, Alice announces the positions of those used for testing, and the participants provide the corresponding measurement results. Alice checks whether any of these outcomes are forbidden for the corresponding encoded states and measurement configurations. If no forbidden outcomes are observed, Alice announces the positions of the rounds retained for secret transmission and her corresponding bit values. Bob and Charlie then privately exchange their measurement results and reconstruct the secret bits using modulo-2 addition of their results and Alice's announced values. After providing an overview of the protocol, we proceed to a detailed analysis of its individual stages.

\subsection{Sharing stage}
At this stage, the generation of the W state is not discussed, since any standard state preparation scheme can be used. Therefore, we proceed directly to the encoding of the secret into the prepared quantum state. The encoding procedure is based on the following reasoning: for all possible outcomes of a Z-basis measurement on the original W state (\ref{W_orig}), the sum of the bits modulo 2 equals 1, while applying an X gate flips the corresponding bit in all possible measurement outcomes, thereby changing the value of this sum. Therefore, the secret bit is encoded according to
\begin{equation}
    s=X_1 \oplus X_2 \oplus X_3 \oplus 1,\quad s, X_1, X_2, X_3 \in \{0,1\}, \label{bit_encode}
\end{equation} 
where $s$ denotes the secret bit, and $X_1$, $X_2$ and $X_3$ are binary indicators specifying whether an X gate is applied to the corresponding qubits. The indices 1, 2 and 3 refer to Alice's, Bob's and Charlie's qubits, respectively, while the state vectors are written in the $CBA$ order.

In the proposed protocol, Alice randomly selects one of two possible combinations of X gates with equal probability for each secret bit. The selection is determined by
\begin{equation}
X_1=1\oplus s,\qquad X_2=X_3=t, \label{x_gates_formula}
\end{equation}
where \(t\in\{0,1\}\) is independently sampled with equal probability for each round, including repeated attempts to transmit the same secret bit. Thus, for $s=0$, Alice applies X gates either to her qubit alone or to all three qubits, while for $s=1$, she applies either no X gates or X gates to both receivers' qubits. The two permitted encoding configurations for each secret bit produce different correlations among the measurement outcomes.

Table~\ref{tab:x_gates} presents the quantum states obtained from the permitted combinations of X gates and the corresponding encoded secret bits.
\renewcommand{\arraystretch}{1.2}

\begin{longtable}{|c|c|c|}
\hline
Applied X gates & State & Secret bit \\
\hline
\endfirsthead

None &
$\frac{1}{\sqrt{3}}(\ket{001}+\ket{010}+\ket{100})$ &
1 \\
\hline

$X_1$ &
$\frac{1}{\sqrt{3}}(\ket{000}+\ket{011}+\ket{101})$ &
0 \\
\hline

$X_2, X_3$ &
$\frac{1}{\sqrt{3}}(\ket{010}+\ket{100}+\ket{111})$ &
1 \\
\hline

$X_1, X_2, X_3$ &
$\frac{1}{\sqrt{3}}(\ket{011}+\ket{101}+\ket{110})$ &
0 \\
\hline

\caption{Quantum states obtained from the permitted combinations of X gates and the corresponding encoded secret bits.}
\label{tab:x_gates}
\end{longtable}

After encoding the secret bit, Alice randomly applies the Hadamard gate to the participants' qubits. Starting from the original W state (\ref{W_orig}), we consider the three configurations in which Alice applies H to Bob's qubit, to Charlie's qubit or to both:
\begin{equation}
    \ket{W_{H_2}}=\frac{1}{\sqrt{6}}(\ket{001}+\ket{011}+\ket{000}-\ket{010}+\ket{100}+\ket{110}), \label{W_h2}
\end{equation}
\begin{equation}
    \ket{W_{H_3}}=\frac{1}{\sqrt{6}}(\ket{001}+\ket{101}+\ket{010}+\ket{110}+\ket{000}-\ket{100}), \label{W_h3}
\end{equation}
\begin{equation}
     \ket{W_{H_{2,3}}}=\frac{1}{2\sqrt{3}}(\ket{001}+\ket{011}+\ket{101}+\ket{111}+2\cdot\ket{000}-2\cdot\ket{110}). \label{W_h23}
\end{equation}
The same approach can be applied to the remaining states in Table~\ref{tab:x_gates}.

\subsection{Reconstruction stage}
After receiving their qubits, Bob and Charlie independently choose between the Z and X measurement bases with equal probability and perform their measurements. They subsequently announce their chosen bases to Alice over an authenticated public channel, while Alice always measures her qubit in the Z basis. Since a measurement in the X basis is equivalent to applying the Hadamard gate before a measurement in the Z basis, applying H by the dealer before a Z-basis measurement by a participant is equivalent to an X-basis measurement without the application of H. Conversely, applying H before an X-basis measurement is equivalent to a Z-basis measurement without the application of H. Therefore, the different combinations of Hadamard-gate applications by the dealer and measurement-basis choices of the participants can be reduced to four distinct effective configurations. These configurations correspond to the cases previously considered for the original W state, namely (\ref{W_orig}), (\ref{W_h2}), (\ref{W_h3}) and (\ref{W_h23}). The remaining variants, arising from different applications of X gates, can be analyzed analogously.

After receiving information about the chosen bases from the participants, Alice takes into account the Hadamard gates she applied and identifies one of the four possible situations. For state (\ref{W_orig}), the participants can uniquely reconstruct the secret bit from their measurement outcomes once the dealer announces her measurement result. In contrast, states (\ref{W_h2}) and (\ref{W_h3}), in which one effective Hadamard gate remains, do not always allow the participants to uniquely determine which secret bit was transmitted, since different possible measurement outcomes may correspond to bit sums of either 0 or 1. Therefore, these cases are used as test cases to detect the presence of an eavesdropper in the channel.

The situation is different for state (\ref{W_h23}). When the dealer's measurement outcome is 0, the corresponding measurement outcomes of the participants can only be 00 or 11, allowing the secret bit to be uniquely determined. Thus, this state can be used for information transmission for this particular measurement outcome of the dealer. When the dealer obtains the other outcome, however, the corresponding representations can be used neither for information transmission nor as test cases, as will be shown in the security analysis. Therefore, we must determine the probability that the round in which this state occurs satisfies the condition for information transmission. For this purpose, we define the corresponding projector:
\begin{equation}
    \Pi_0=\mathbb{I}\ \otimes\ \mathbb{I}\ \otimes\ \ketbra{0}{0}, \label{P_0}
\end{equation}
applying it to (\ref{W_h23}), we obtain
\begin{equation}
    \Pi_0\ket{W_{H_{2,3}}}=\frac{1}{2\sqrt{3}}(2\cdot\ket{000}-2\cdot\ket{110}). \label{P_0W_h23}
\end{equation}
Let $q_A$ denote Alice's raw measurement outcome. For the state (\ref{W_h23}), the probability of obtaining $q_A=0$ is
\begin{equation}
    P(q_A=0)=\left\|\Pi_0 \ket{W_{H_{2,3}}}\right\|^2=\frac{2}{3}. \label{P_meas_0}
\end{equation}
When the X gates are applied to the participants' qubits, the same reasoning applies, since the situation with respect to the dealer remains unchanged. However, when the X gate is applied to the dealer's qubit, a different projector must be considered:
\begin{equation}
    \Pi_1=\mathbb{I}\ \otimes\ \mathbb{I}\ \otimes\ \ketbra{1}{1}, \label{P_1}
\end{equation}
applying it, we obtain
\begin{equation}
    \Pi_1 (\mathbb{I}\ \otimes\ \mathbb{I}\ \otimes\ \mathbb{X})\ket{W_{H_{2,3}}}=\frac{1}{2\sqrt{3}}(2\cdot\ket{001}-2\cdot\ket{111}), \label{P_1W_h23}
\end{equation}
which gives the same probability
\begin{equation}
    P(q_A=1)=\left\|\Pi_1 (\mathbb{I}\ \otimes\ \mathbb{I}\ \otimes\ \mathbb{X})\ket{W_{H_{2,3}}}\right\|^2=\frac{2}{3}. \label{P_meas_1}
\end{equation}
Therefore, for state (\ref{W_h23}), as well as its variants obtained by the permitted combinations of X gates, the probability that a semi-information round satisfies the condition for information transmission is $2/3$.

In the ideal no-attack model, the suitable semi-information outcomes give equal measurement results for Bob and Charlie, while Alice's raw result is $1-s$, as follows from (\ref{P_0W_h23}) and (\ref{P_1W_h23}). The modulo-2 sum of the three raw results therefore equals $1-s$ instead of the encoded bit $s$. Alice consequently inverts her result before announcing it, allowing the same modulo-2 reconstruction rule to be used for information and semi-information rounds. Let $M_A$ denote her announced value, including this inversion. In a semi-information round used for transmission, this gives $M_A=s$.

In an ordinary information round, $P(M_A=s)=1/3$. Information and semi-information configurations are equally likely, as summarized in Table~\ref{tab:rounds}, but only $2/3$ of semi-information rounds satisfy the condition for information transmission. If all suitable semi-information rounds were used, ordinary information rounds would therefore constitute $3/5$ of the rounds used for secret transmission, while suitable semi-information rounds would constitute $2/5$. Since $M_A=s$ in every suitable semi-information round used for transmission, the resulting probability would be $P(M_A=s\mid\text{retained})=(3/5)(1/3)+2/5=3/5$. An observer could then guess the secret bit correctly with probability $3/5$ simply by taking Alice's announced value as the guess.

To remove this advantage, Alice additionally selects each suitable semi-information round for information transmission with probability $1/2$, independently of the encoded secret bit and the applied X-gate combination. Otherwise, the round is discarded. The probability that a semi-information round is used for information transmission is therefore $(2/3)(1/2)=1/3$. Ordinary information rounds and retained semi-information rounds then constitute fractions $3/4$ and $1/4$ of the retained rounds, respectively, giving
\[
P(M_A=s\mid\text{retained})=\frac34\cdot\frac13+\frac14=\frac12.
\]
Thus, Alice's announced value is equally likely to be 0 or 1 for either secret-bit value and reveals no information about the secret on its own. Alice announces the retained positions as a common list, without identifying the semi-information rounds or revealing her corresponding Hadamard-gate choices, since identifying a retained semi-information round would disclose its secret bit directly through $M_A=s$.

Table~\ref{tab:rounds} provides a summary of the considered configurations, showing the Hadamard gates applied by Alice to the transmitted qubits, the measurement bases chosen by Bob and Charlie and the type of the round:
\begin{longtable}{|c|c|c|c|c|}
\hline
\makecell{Alice's preparation / \\ Participants' measurement} &
$Z, Z$ &
$X, Z$ &
$Z, X$ &
$X, X$ \\
\hline
\endfirsthead

$I, I$ &
I &
T &
T &
S-I \\
\hline

$H, I$ &
T &
I &
S-I &
T \\
\hline

$I, H$ &
T &
S-I &
I &
T \\
\hline

$H, H$ &
S-I &
T &
T &
I \\
\hline

\caption{Summary of the Hadamard gates applied by Alice, participants' measurement bases and the corresponding round types.}
\label{tab:rounds}
\end{longtable}
\noindent
The rows correspond to the different configurations of Hadamard gates applied by Alice to the qubits sent to Bob and Charlie, while the columns correspond to the measurement bases chosen by the participants. Each cell indicates the corresponding round type: "I" denotes an information round, "T" denotes a test round and "S-I" denotes a semi-information round that can be used for information transmission only for certain measurement outcomes of the dealer. The notation in the rows and columns follows the same order for Bob and Charlie: the first entry refers to Bob's qubit, and the second entry refers to Charlie's qubit. Thus, for example, $H,I$ in a row indicates that Alice applies the Hadamard gate to Bob's qubit and does not apply it to Charlie's qubit, while $X,Z$ in a column indicates that Bob measures his qubit in the X basis and Charlie measures his qubit in the Z basis.

After performing the measurements, all participants record their results. The protocol continues until a sufficient number of information rounds, including the selected semi-information rounds, has been obtained to transmit the entire secret sequence. If a round is classified as a test round or is discarded, Alice repeats the encoding of the current secret bit in a newly prepared state and proceeds to the next secret bit only after a round has been retained for transmission. It should be noted that the total number of rounds is not fixed in advance, since the occurrence of information rounds depends on the random choices of measurement bases, Hadamard-gate applications and the additional selection of semi-information rounds. At the same time, the dealer knows when the last information bit has been transmitted and announces the end of the transmission at that point. 

The dealer then announces the positions of the test rounds, and the participants report their measurement results for these rounds. The dealer examines the test results for outcomes that are forbidden for the corresponding encoded states and measurement configurations. If any such outcome is observed, the protocol is terminated. Otherwise, the dealer announces the positions of the rounds retained for transmission together with her corresponding bit values. Semi-information rounds that do not satisfy the required measurement condition or are not selected by Alice are discarded and are not included among the announced information rounds. Bob and Charlie then privately exchange their measurement results for the retained rounds and recover each secret bit as follows:
\begin{equation}
    s_r=M_A \oplus M_B \oplus M_C,\quad s_r, M_A, M_B, M_C \in \{0,1\}, \label{bit_decode}
\end{equation}
where $s_r$ denotes the recovered bit of the secret sequence and $M_B$ and $M_C$ denote the measurement results of Bob and Charlie, respectively.

\section{Efficiency and security}\label{sec:3}

In this section, we analyze the efficiency of the proposed protocol and evaluate its security against two attack models: intercept-and-resend and entangle-and-measure. The attacks are considered in the presence of an internal adversary, since such an adversary has more information than an external adversary and therefore poses a more relevant threat. We use the original W state (\ref{W_orig}) to present the test-round outcome distributions arising from Alice's Hadamard-gate applications and the measurement bases chosen by Bob and Charlie. The permitted X-gate combinations may change which outcomes are forbidden. For the intercept-and-resend attack considered below, they do not change the total forbidden-outcome probability. In the entangle-and-measure analysis, we take all permitted encodings into account when restricting Bob's interaction. For the resulting class of interactions, the total forbidden-outcome probability is likewise independent of the permitted X-gate combination, so we present the calculation using the original W state. For convenience, we assume throughout the following analysis that Bob is the dishonest participant, while the same reasoning applies if Charlie is dishonest. In both attack models, Bob intervenes on Charlie's transmitted qubit before the measurement bases are announced.

\subsection{Analysis of efficiency}
We now evaluate the efficiency of the protocol by determining the expected number of qubits required to transmit a secret of length $n$. Let $R=\{R_1,R_2,\dots,R_m\}$ denote the set of rounds prepared by the dealer, indexed in their order of transmission, where $m$ is the total number of rounds. Among these rounds, let $I\subseteq R$ denote the set of rounds retained for information transmission and $T\subseteq R$ the set of test rounds, with $I\cap T=\varnothing$. The remaining rounds are discarded and are used neither for information transmission nor for security testing. 

Let us determine the probability that a given round $R_i \in R$ belongs to $I$. As summarized in Table~\ref{tab:rounds}, there are 16 possible combinations of Hadamard-gate applications by Alice and measurement bases chosen by Bob and Charlie, corresponding to three types of rounds: information rounds, test rounds and semi-information rounds. Each combination occurs with equal probability, since Alice's Hadamard-gate applications and the participants' choices of measurement bases are made independently, with each option having probability 1/2. Therefore, the probability of obtaining a round of type $I$ in Table~\ref{tab:rounds} is 1/4, corresponding to the four configurations in which the applications of Hadamard gates by Alice are consistent with the measurement bases chosen by Bob and Charlie. The configurations corresponding to the semi-information rounds should also be taken into account. Their total probability is 1/4, while the probability that such a configuration satisfies the condition for information transmission is 2/3, as follows from (\ref{P_meas_0}) and (\ref{P_meas_1}). However, Alice additionally selects each suitable semi-information round for information transmission with probability 1/2. Consequently, the overall probability $P(R_i \in I)$ is
\begin{equation}
    P(R_i \in I)= \frac{1}{4}+\frac{1}{4} \cdot \frac{2}{3} \cdot \frac{1}{2}=\frac{1}{3}. \label{Prob_inf}
\end{equation}
Thus, on average, 1/3 of the transmitted rounds are used for information transmission. Since each information round contributes one bit to the secret sequence, the expected number of transmitted rounds required to share a secret of length $n$ is
\begin{equation}
    E[m]= \frac{n}{P(R_i \in I)}=\frac{n}{1/3}=3n. \label{E_m}
\end{equation}
Since each round requires a three-qubit W state, including the qubit retained by Alice, the expected total number of qubits used is
\begin{equation}
    3E[m]=3\cdot3n=9n. \label{3E_m}
\end{equation}

To evaluate the quantum resource consumption of the protocol, we define the qubit efficiency as the ratio of the secret length to the expected total number of prepared qubits, including Alice's retained qubits. Using the previously derived value of $E[m]$, we obtain
\begin{equation}
    \eta_q=\frac{n}{3E[m]}=\frac{n}{9n}=\frac{1}{9}\approx11.11\%. \label{eff}
\end{equation}

\subsection{Intercept-and-resend attack}
We now consider an internal intercept-and-resend attack performed by Bob, who attempts to obtain information about the secret without the assistance of Charlie. In addition to his own qubit, Bob intercepts the qubit transmitted to Charlie, measures it in the Z or X basis with equal probability and subsequently resends a new qubit to Charlie according to the measurement outcome.

For a single qubit, which in our case is Charlie's qubit, we consider the following cases: Alice either applies or does not apply the H gate, Bob measures the intercepted qubit in the Z or X basis, and Charlie subsequently measures it in the Z or X basis. Therefore, there are $2\cdot2\cdot2=8$ possible combinations. However, not all of these combinations need to be analyzed separately, since some do not introduce detectable changes in the test statistics.

First, when Bob and Charlie measure Charlie's qubit in the same basis, Bob's intervention does not alter the joint measurement statistics. In this case, Bob effectively performs Charlie's measurement in advance and resends the corresponding eigenstate, so the final outcome distribution remains unchanged.

Now consider the cases in which Bob and Charlie choose different measurement bases. When Bob measures in the Z basis without Alice applying H to Charlie's qubit, or in the X basis after Alice applies H, the corresponding test configurations remain statistically unaffected by the attack. This follows from the correlations of the W state: conditioning on Alice's and Bob's effective Z-basis measurement outcomes on their respective qubits leaves Charlie's qubit in a definite state, either $\ket{0}$ or $\ket{1}$, before any Hadamard gate is applied to it. Applying H to Charlie's qubit transforms the corresponding states $\ket{0}$ and $\ket{1}$ into $\ket{+}$ and $\ket{-}$. Therefore, Bob's measurement of the intercepted qubit in the respective basis preserves these correlations, and Charlie's subsequent measurement produces the same joint outcome distribution as in the absence of the attack. These cases correspond to the test configuration (\ref{W_h3}).

Consequently, the attack can be detected in the remaining two cases: when Alice does not apply H to Charlie's qubit, Bob measures it in the X basis and Charlie measures it in the Z basis, or when Alice applies H to Charlie's qubit, Bob measures it in the Z basis and Charlie measures it in the X basis. In both cases, the test configuration (\ref{W_h2}) is suitable for detecting the adversary. We therefore focus on these two cases in the following analysis.

We take state (\ref{W_h2}) as the reference state corresponding to the absence of an adversary in the channel. For this state, only the following outcomes are possible, with the corresponding probabilities:
\begin{equation}
    P(000)=P(001)=P(010)=P(011)=P(100)=P(110)=\frac{1}{6}. \label{pos}
\end{equation}
The idea behind detecting the adversary is that his presence introduces a nonzero probability of measurement outcomes that are impossible in the absence of an adversary. In this case, these outcomes are
\begin{equation}
    P(101)=P(111)=0. \label{impos}
\end{equation}

We begin the analysis of the intervention with the first case, examining how the adversary modifies state (\ref{W_h2}). Since Bob measures Charlie's qubit in the X basis, the two possible measurement outcomes, + and -, occur with equal probability of 1/2. For the outcome +, the resulting state of the three-qubit system is
\begin{equation}
    \ket{\psi_+}=\ket{+}\otimes\frac{1}{\sqrt{6}}(2\ket{00}+\ket{01}+\ket{11}). 
\end{equation}
Accordingly, the qubit resent by Bob to Charlie is in the state $\ket{+}$. Charlie subsequently measures the received qubit in the Z basis. For the measurement outcome 0, the resulting state of the three-qubit system is
\begin{equation}
    \ket{\phi_+,0}=\frac{1}{\sqrt{6}}(2\ket{000}+\ket{001}+\ket{011}), \label{+_0}
\end{equation}
while for the measurement outcome 1, the resulting state is
\begin{equation}
    \ket{\phi_+,1}=\frac{1}{\sqrt{6}}(2\ket{100}+\ket{101}+\ket{111}). \label{+_1}
\end{equation}
Since Bob obtains the outcome + with probability 1/2 and each outcome of Charlie's Z-basis measurement occurs with conditional probability 1/2, the two branches (\ref{+_0}) and (\ref{+_1}) each occur with probability 1/4.

For the outcome -, the resulting state of the three-qubit system is
\begin{equation}
    \ket{\psi_-}=\ket{-}\otimes\frac{1}{\sqrt{6}}(\ket{01}-2\ket{10}+\ket{11}). \label{34}
\end{equation}
In this case, the qubit resent by Bob to Charlie is in the state $\ket{-}$. Following Charlie's Z-basis measurement, the two possible resulting states are
\begin{equation}
    \ket{\phi_-,0}=\frac{1}{\sqrt{6}}(\ket{001}-2\ket{010}+\ket{011}), \label{-_0}
\end{equation}
and
\begin{equation}
    \ket{\phi_-,1}=\frac{1}{\sqrt{6}}(\ket{101}-2\ket{110}+\ket{111}). \label{-_1}
\end{equation}
Both resulting states (\ref{-_0}) and (\ref{-_1}) occur with probability 1/4 each, as in the previous case.

The probabilities of the possible outcomes obtained by combining all four branches yield the following distribution:
\begin{align}
    P(000)=P(010)=P(100)=P(110)&=\:\frac{1}{6}; \\
    P(001)=P(011)=P(101)=P(111)&=\frac{1}{12}. \label{prob_impos}
\end{align}
As can be seen, the probability distribution has changed, and the previously forbidden measurement outcomes from (\ref{impos}) now occur with nonzero probability, as shown in (\ref{prob_impos}). Hence, the probability of detecting the attack in a single test round is
\begin{equation}
     P_{\mathrm{det}\mid\mathrm{test}}=P(101)+P(111)=\frac{1}{6}.\label{prob_det}
\end{equation}

The second case is analyzed similarly, where Alice applies the H gate to Charlie's qubit, after which Bob intercepts the qubit, measures it in the Z basis and resends it to Charlie, who measures it in the X basis. The resulting probability distribution will be exactly the same, and the probability of detecting the attack in a single round is also 1/6.

Now consider why semi-information rounds that do not satisfy the condition for information transmission cannot be used as test rounds either, taking (\ref{W_h23}) as a reference state corresponding to the absence of an adversary. This state can arise in four cases, depending on the Hadamard gates applied by Alice and the measurement bases chosen by Bob and Charlie. However, the way in which the effective H operation on Bob's qubit is obtained does not affect the following analysis. Therefore, we consider two cases: when Alice applies H to Charlie's qubit and Charlie measures it in the Z basis, and when Alice does not apply H to Charlie's qubit and Charlie measures it in the X basis. As an example, consider the first case, for which it can be shown that, regardless of the measurement basis chosen by Bob upon interception, the resulting probability distribution of the measurement outcomes remains unchanged and Bob's intervention cannot be detected statistically.

For state (\ref{W_h23}), obtaining the result 1 by the dealer means that the given round cannot be used to transmit information. The state of Bob and Charlie then reduces to
\begin{equation}
    \ket{\psi_{BC}}=\frac{1}{2}(\ket{00}+\ket{01}+\ket{10}+\ket{11})=\ket{++}.\label{psi_BC}
\end{equation}
Charlie measures his qubit in the Z basis. Hence, in the absence of an adversary, the only two possible outcomes are
\begin{equation}
     P_{\mathrm{ref}}(C=0)=P_{\mathrm{ref}}(C=1)=\frac{1}{2}.
\end{equation}
If Bob measures Charlie's qubit in the Z basis, the same two possible outcomes occur:
\begin{equation}
     P(B_{\mathrm{int}}=0)=P(B_{\mathrm{int}}=1)=\frac{1}{2}.
\end{equation}
Bob then forwards the corresponding state, $\ket{0}$ or $\ket{1}$, to Charlie. Consequently, the probability of Charlie obtaining either 0 or 1 remains unchanged in the presence of Bob's intervention, being equal for both outcomes. Therefore, the intervention cannot be detected statistically. When Bob measures Charlie's qubit in the X basis, the $\ket{+}$ state is obtained with probability 1, as shown by (\ref{psi_BC}). Bob then forwards this state to Charlie without modification, just as it would be transmitted in the absence of his intervention. Therefore, his intervention remains statistically undetected in this case as well.

The second case, in which Alice does not apply H to Charlie's qubit and Charlie measures it in the X basis, is analyzed analogously. In this case, Bob also remains statistically undetected regardless of the measurement basis he chooses for the intercepted qubit. Thus, semi-information rounds that do not satisfy the condition for information transmission are not used as test rounds either.

Based on the results obtained for the individual cases, the overall probability of detecting the adversary can now be determined. Four possible configurations are considered: (\ref{W_orig}), (\ref{W_h2}), (\ref{W_h3}) and (\ref{W_h23}). In the case of Bob acting as the adversary, only configuration (\ref{W_h2}) can be used for detection. As Alice's Hadamard-gate applications and the participants' measurement-basis choices are made independently, with each option having probability 1/2, the probability of obtaining configuration (\ref{W_h2}) in a given round is 1/4. However, the attack can be detected only when Bob selects the corresponding incompatible measurement basis upon interception, which occurs with probability 1/2. Therefore, using (\ref{prob_det}), the overall probability of detecting the adversary in a single round is
\begin{equation}
     P_{\mathrm{det}}=P_{\mathrm{test}} \cdot P_{\mathrm{det}\mid\mathrm{test}}=\frac{1}{4}\cdot\frac{1}{2}\cdot\frac{1}{6}=\frac{1}{48}.\label{prob_det_gen}
\end{equation}
Since a single forbidden outcome in a test round is sufficient to detect the adversary, we can determine the cumulative detection probability before the transmission of the entire secret sequence is completed. Retaining a round for information transmission and detecting the attack are mutually exclusive events. Accounting for attempts in which neither event occurs, the probability that the next occurrence of either event is a retained round is $P(R_i\in I)/(P(R_i\in I)+P_{\mathrm{det}})$. Assuming that the attack is performed independently and identically in each round, the probability that at least one forbidden outcome occurs before $n$ rounds have been retained for secret transmission is given by
\begin{equation}
    P_{\mathrm{det}}^{(n)}=1-\bigg(\frac{P(R_i\in I)}{P(R_i\in I)+P_{\mathrm{det}}}\bigg)^n=1-\bigg(\frac{16}{17}\bigg)^n. \label{prob_det_ir_cumulative}
\end{equation}
For example, for a 256-bit secret, the cumulative detection probability is approximately $0.99999982$, indicating that the considered attack can be detected with high probability when the test results are subsequently examined.
\subsection{Entangle-and-measure attack}
We next consider an entangle-and-measure attack performed by Bob on Charlie's qubit. In contrast to the intercept-and-resend attack, Bob does not need to measure the intercepted qubit immediately. Instead, he may entangle it with an ancillary system $E$, retain the ancillary system and postpone its measurement. The interaction between Charlie's intercepted qubit and the ancillary system $E$ is described by the unitary operator $U_{CE}$. The subsequent analysis focuses on the information that Bob can obtain through this interaction with the intercepted qubit, without considering joint operations involving his own qubit.

For this attack, we focus on two types of test cases corresponding to configuration (\ref{W_h2}): when Alice does not apply H to Charlie's qubit and Charlie measures it in the Z basis, or when Alice applies H to Charlie's qubit and Charlie measures it in the X basis. In both cases, the effective H operation on Bob's qubit can arise either from Alice applying H followed by Bob's Z-basis measurement or from Bob's X-basis measurement without Alice applying H. As established in the intercept-and-resend analysis, configuration (\ref{W_h3}) does not allow the detection of certain measurement-based attacks. Similarly, under an entangle-and-measure attack, Bob can choose an interaction that allows him to obtain information about the intercepted qubit while preserving the measurement statistics of this configuration. Therefore, we restrict the following analysis to configuration (\ref{W_h2}), which allows us to establish the conditions under which Bob's intervention can remain undetected.

Suppose that the initial state of Bob's ancillary system is $\ket{E_0}$. In the most general case, the interaction $U_{CE}$ can be expressed as follows:
\begin{align}
    U_{CE}\ket{0}_C\ket{E_0}_E&=\ket{0}_C\ket{e_0}_E+\ket{1}_C\ket{e_1}_E,\\
    U_{CE}\ket{1}_C\ket{E_0}_E&=\ket{0}_C\ket{e_2}_E+\ket{1}_C\ket{e_3}_E.
\end{align}
Here, $\ket{e_i}$ are generally unnormalized vectors of the ancillary system that include the corresponding transition amplitudes. Unitarity requires the two joint output states to be normalized and mutually orthogonal:
\begin{equation}
\begin{gathered}
\|\ket{e_0}\|^2+\|\ket{e_1}\|^2=1,\qquad
\|\ket{e_2}\|^2+\|\ket{e_3}\|^2=1,\\
\braket{e_0}{e_2}+\braket{e_1}{e_3}=0.
\end{gathered}
\end{equation}
The terms containing $\ket{e_1}$ and $\ket{e_2}$ correspond to the transitions:
\begin{equation}
    \ket{0}_C\xrightarrow{}\ket{1}_C,\quad  \ket{1}_C\xrightarrow{}\ket{0}_C, \label{trans}
\end{equation}
while $\ket{e_0}$ and $\ket{e_3}$ correspond to Charlie's qubit state being preserved.

For the subsequent analysis, we first consider the case in which Charlie measures his qubit in the Z basis and Alice does not apply H to it. In the absence of an adversary, the allowed measurement outcomes are determined by (\ref{pos}) and the applied X-gate combination. Hence, depending on the applied X-gate combination, the transitions in (\ref{trans}) lead to forbidden measurement outcomes. To avoid producing such errors, Bob's interaction must satisfy
\begin{equation}
    \ket{e_1}=0, \quad \ket{e_2}=0.
\end{equation}
Thus, the attacks that remain indistinguishable from the ideal behavior of the protocol are limited to interactions of the following form:
\begin{align}
    \ket{0}_C\ket{E_0}_E&\xrightarrow{}\ket{0}_C\ket{e_0}_E,\\
    \ket{1}_C\ket{E_0}_E&\xrightarrow{}\ket{1}_C\ket{e_3}_E.
\end{align}
Consequently, Bob can potentially obtain information about Charlie's qubit through the ancillary system without changing the two states considered above. This is possible when the corresponding ancillary states are physically distinguishable, that is, when
\begin{equation}
\left|\braket{e_0}{e_3}\right|<1.
\end{equation}
However, ancillary states differing only by a phase factor are physically indistinguishable and therefore cannot provide Bob with information about Charlie's qubit.

Now consider the effect of this attack in the test configuration in which Alice applies H to Charlie's qubit and Charlie measures his qubit in the X basis. Before the first Hadamard gate is applied, consider, for example, Charlie's qubit in the state $\ket{0}_C$, since the same reasoning applies when its initial state is $\ket{1}_C$. After the first H gate, its state becomes $\ket{+}_C=\frac{\ket{0}_C+\ket{1}_C}{\sqrt{2}}$. The subsequent interaction with the ancillary system gives

\begin{equation}
    \ket{+}_C\ket{E_0}_E\xrightarrow{}\frac{1}{\sqrt{2}} (\ket{0}_C\ket{e_0}_E+\ket{1}_C\ket{e_3}_E).
\end{equation}

After this, Charlie performs an X-basis measurement. This measurement is implemented by applying the H gate to his qubit before the measurement itself. Thus, the action of the second H gate results in
\begin{equation}
    \frac{1}{2}(\ket{0}_C(\ket{e_0}_E+\ket{e_3}_E)+\ket{1}_C(\ket{e_0}_E-\ket{e_3}_E)).
\end{equation}
In the absence of Bob's intervention, Charlie always obtains the measurement outcome 0, and therefore the outcome 1 is impossible. Under the entangle-and-measure attack, however, the probability of obtaining this outcome becomes
\begin{equation}
    P_{\mathrm{err}}=\frac{1}{4}\|\ket{e_0}-\ket{e_3}\|^2. \label{prob_em}
\end{equation}
Therefore, any distinguishability between the corresponding ancillary states, which is necessary for Bob to obtain information about Charlie's qubit through the ancillary system, leads to a deviation from the reference statistics.

The probabilities of the impossible outcomes in (\ref{impos}) can now be determined for this attack. The outcome 101 may arise as an erroneous result from 001, and similarly 111 may arise from 011. In the following expressions, $P(001)$ and $P(011)$ refer to the reference distribution without the attack, where both outcomes occur with probability 1/6, whereas $P(101)$ and $P(111)$ refer to the distribution under the attack. Therefore, taking (\ref{prob_em}) into account, the probability of each of these impossible outcomes is
\begin{align}
    P(101)&=P(001)\cdot P_{\mathrm{err}}=\frac{1}{6}\cdot \frac{1}{4}\|\ket{e_0}-\ket{e_3}\|^2=\frac{1}{24}\|\ket{e_0}-\ket{e_3}\|^2,\\
    P(111)&=P(011)\cdot P_{\mathrm{err}}=\frac{1}{6}\cdot \frac{1}{4}\|\ket{e_0}-\ket{e_3}\|^2=\frac{1}{24}\|\ket{e_0}-\ket{e_3}\|^2.
\end{align}
Thus, the total probability of obtaining a forbidden outcome in a single test round is
\begin{equation}
     P_{\mathrm{det}\mid\mathrm{test}}=P(101)+P(111)=\frac{1}{12}\|\ket{e_0}-\ket{e_3}\|^2. \label{prob_det_em}
\end{equation}
It follows that
\begin{equation}
    P_{\mathrm{det}\mid\mathrm{test}}=0 \Longleftrightarrow \ket{e_0}=\ket{e_3}, 
\end{equation}
whereas
\begin{equation}
    \ket{e_0}\neq\ket{e_3} \Longleftrightarrow P_{\mathrm{det}\mid\mathrm{test}}>0.
\end{equation}
Consequently, Bob faces an information-disturbance trade-off: obtaining information about Charlie's state through the ancillary system requires the corresponding ancillary states to be physically distinguishable, which implies $\ket{e_0}\neq\ket{e_3}$ and therefore leads to a nonzero probability of forbidden outcomes. Complete absence of detectable disturbance requires $\ket{e_0}=\ket{e_3}$, so the interaction then becomes
\begin{equation}
    \ket{0}_C\ket{E_0}_E\xrightarrow{}\ket{0}_C\ket{e}_E, \quad \ket{1}_C\ket{E_0}_E\xrightarrow{}\ket{1}_C\ket{e}_E,
\end{equation}
giving
\begin{equation}
    \rho_E^{(0)} = \rho_E^{(1)} = \ketbra{e}{e}. 
\end{equation}
Hence, measuring the ancilla does not allow Bob to distinguish between 0 and 1 on Charlie's qubit.

Analogously to the previous attack, we determine the overall probability of detecting the adversary under the entangle-and-measure attack. For the entangle-and-measure interactions satisfying $\ket{e_1}=\ket{e_2}=0$, only half of the test cases corresponding to configuration (\ref{W_h2}) allow the detection of Bob's intervention. Therefore, the required test cases occur with probability 1/8. The conditional detection probability is given in (\ref{prob_det_em}), and hence the overall detection probability is
\begin{equation}
     P_{\mathrm{det}}=P_{\mathrm{test}} \cdot P_{\mathrm{det}\mid\mathrm{test}}=\frac{1}{8}\cdot\frac{1}{12}\|\ket{e_0}-\ket{e_3}\|^2=\frac{1}{96}\|\ket{e_0}-\ket{e_3}\|^2.\label{prob_det_em_gen}
\end{equation}
Assuming that Bob applies the same interaction independently in each round, the probability that at least one forbidden outcome occurs before $n$ rounds have been retained for secret transmission is given by
\begin{equation}
    P_{\mathrm{det}}^{(n)}=1-\bigg(\frac{32}{32+\|\ket{e_0}-\ket{e_3}\|^2}\bigg)^n. \label{prob_det_em_cumulative}
\end{equation}

We now quantify the information that Bob can obtain about Charlie's qubit through his ancillary system for the interactions satisfying the previously derived conditions $\ket{e_1}=\ket{e_2}=0$. In particular, we evaluate his ability to distinguish the corresponding ancillary states, which is distinct from obtaining information about the encoded secret bit.

For the considered interactions, the states $\ket{0}_C$ and $\ket{1}_C$ are associated with the normalized ancillary states $\ket{e_0}$ and $\ket{e_3}$, respectively. Their corresponding density operators are
\begin{equation}
\rho_E^{(0)}=\ketbra{e_0}{e_0},\qquad \rho_E^{(1)}=\ketbra{e_3}{e_3}.
\end{equation}
The distinguishability of these states is characterized by their trace distance:
\begin{equation}
D=\frac{1}{2}\left\|\rho_E^{(0)}-\rho_E^{(1)}\right\|_1=\sqrt{1-\left|\braket{e_0}{e_3}\right|^2}.
\end{equation}
Since the two ancillary states occur with equal probability due to Alice's random encoding, the optimal probability of correctly distinguishing them is given by the Helstrom bound:
\begin{equation}
P_{\mathrm{guess}}=\frac{1+D}{2}.
\end{equation}
The corresponding minimum error probability is
\begin{equation}
1-P_{\mathrm{guess}}=\frac{1-D}{2}.
\end{equation}
For the two equiprobable pure ancillary states considered here, an optimal minimum-error measurement also maximizes the mutual information \cite{Ban1997}. The accessible information is therefore
\begin{equation}
I_{\mathrm{acc}}=1-h_2\left(\frac{1-D}{2}\right),
\end{equation}
where
\begin{equation}
h_2(p)=-p\log_2p-(1-p)\log_2(1-p)
\end{equation}
denotes the binary Shannon entropy. Here, $I_{\mathrm{acc}}$ characterizes the maximum mutual information that Bob can obtain by measuring the ancillary system alone, without including additional information from his own qubit or subsequent classical announcements.

Finally, using the previously derived detection probability and the normalization of the ancillary states, we obtain
\begin{equation}
P_{\mathrm{det}}=\frac{1-\operatorname{Re}\braket{e_0}{e_3}}{48}.
\end{equation}
Since
\begin{equation}
\operatorname{Re}\braket{e_0}{e_3}\leq\left|\braket{e_0}{e_3}\right|=\sqrt{1-D^2},
\end{equation}
it follows that
\begin{equation}
P_{\mathrm{det}}\geq\frac{1-\sqrt{1-D^2}}{48}.
\end{equation}
Together with the expression for $I_{\mathrm{acc}}$, this establishes a quantitative trade-off between information gain and disturbance. Increasing the distinguishability of the ancillary states increases both the accessible information and the lower bound on the detection probability. The bound is attained when $\braket{e_0}{e_3}$ is real and nonnegative.

\section{Software implementation and results}
\label{sec:4}

This section presents the software implementation of the proposed protocol in Python using the Qiskit framework \cite{QISkit} and its evaluation using the Qiskit Aer simulator and an IBM Quantum processor. One protocol round is constructed as described below.

The process starts with all qubits initialized in the state $\ket{0}$. The W state is then generated using the circuit proposed in \cite{Diogo2019}. Next, Alice randomly applies X gates according to the configurations specified in Table~\ref{tab:x_gates} to encode the secret bit and then applies a Hadamard gate to Bob's and to Charlie's qubit independently, each with probability $1/2$.

In the actual protocol, Alice then sends the corresponding qubits to Bob and Charlie. In the software implementation, this step is represented by assigning qubits 1 and 2 to Bob and Charlie, respectively, while qubit 0 remains Alice's qubit. Thus, software qubits 0, 1 and 2 correspond to A, B and C and to positions 1, 2 and 3 used for the X-gate indicators in (\ref{bit_encode}) and (\ref{x_gates_formula}).

\begin{figure}[htbp]
    \centering
    \includegraphics[width=0.8\linewidth]{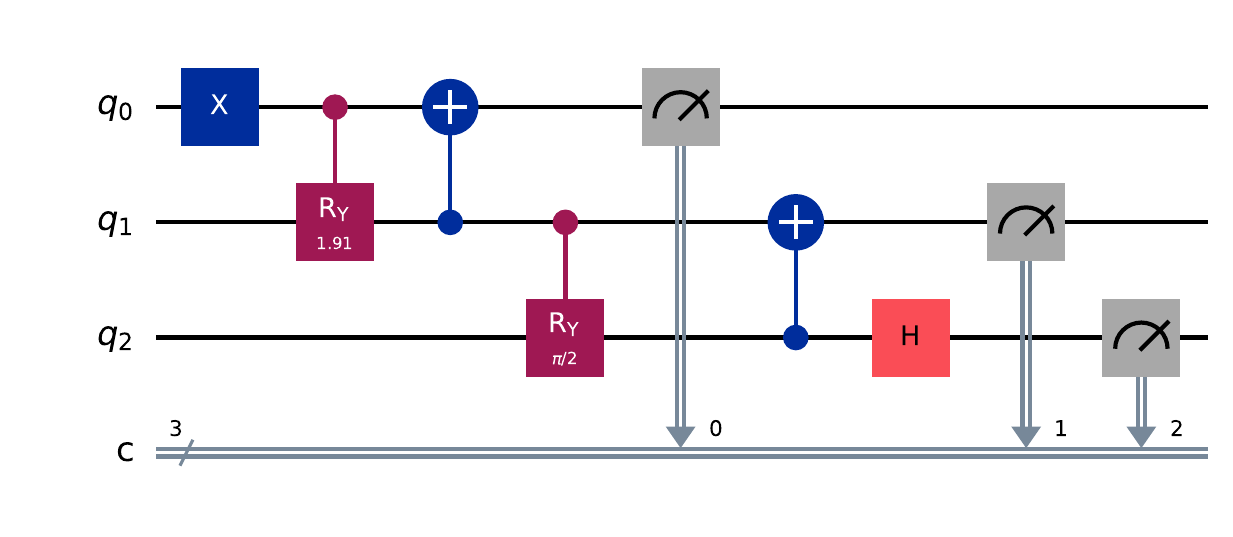}
    \caption{Circuit representation of a complete protocol round.}
    \label{round}
\end{figure}

Alice measures her qubit in the Z basis, whereas Bob and Charlie independently choose between the Z and X bases with equal probability. A Z-basis measurement is performed directly, while an X-basis measurement is implemented by applying a Hadamard gate before measurement. Finally, the round type is determined from Alice's Hadamard-gate applications and the measurement bases chosen by Bob and Charlie, according to Table~\ref{tab:rounds}. Together, these steps constitute one complete protocol round. Figure~\ref{round} shows an example of the circuit corresponding to such a round.

The round shown in Figure~\ref{round} is classified as a test round, since one effective Hadamard gate remains on Charlie's qubit, corresponding to (\ref{W_h3}).

The program constructs subsequent rounds according to the same procedure. If a round is used for testing or discarded, Alice repeats the encoding of the current secret bit in a newly prepared state. The program proceeds to the next secret bit only after an information round or a retained semi-information round has been obtained. Suitable semi-information rounds are retained with the additional probability $1/2$ specified in Section~\ref{sec:2}.

In the ideal illustrative implementation, each test round is checked immediately after measurement to avoid unnecessary execution. If a forbidden outcome is detected, the program terminates without releasing a recovered secret. This implements the same accept-or-abort rule as checking all test outcomes at the end of transmission. If all tests pass, the secret is reconstructed from information rounds and retained semi-information rounds. For each retained semi-information round, Alice inverts her raw measurement result before announcing it, and the secret bit is then reconstructed using the rule (\ref{bit_decode}). The statistical criterion used to analyze hardware measurements is described separately in Subsection~\ref{sec:4_statistical}.

For clarity, the following output shows the choices and outcomes recorded during one run on the Qiskit Aer simulator. The input secret is $01001$. The output was obtained by executing the protocol with fresh random choices.

\begin{lstlisting}[basicstyle=\ttfamily\footnotesize,breaklines=true]
X:          X0, X1, X2 | I    | I    | X0, X1, X2 | X0     | X0, X1, X2 | X0   | I   
H:          I          | I    | H1   | I          | H1, H2 | I          | H2   | I   
Basis:      X, X       | Z, X | X, Z | Z, Z       | Z, Z   | Z, X       | X, Z | X, X
Round:      S-I        | T    | I    | I          | S-I*   | T          | S-I  | S-I*
Outcome:    111        | 110  | 001  | 011        | 110    | 011        | 111  | 110 

X:          X1, X2 | X1, X2
H:          H1, H2 | H2    
Basis:      Z, X   | Z, X  
Round:      T      | I     
Outcome:    111    | 010   
Recovered:  0 1 0 0 1
Secret:     0 1 0 0 1
\end{lstlisting}

The symbol $|$ separates entries belonging to consecutive rounds. In the \texttt{X} and \texttt{H} rows, I denotes that no gate was applied at the corresponding stage, while the numbers identify the software qubits 0, 1 and 2. The \texttt{X} row specifies Alice's X-gate choices for encoding, and the \texttt{H} row specifies her Hadamard-gate applications. Each entry in the \texttt{Basis} row contains Bob's measurement basis followed by Charlie's.

The \texttt{Round} row gives the round type using the notation of Table~\ref{tab:rounds}; S-I* denotes a semi-information round that is discarded because it fails the dealer's measurement condition or is not selected for retention.

The \texttt{Outcome} row contains the raw measurement results in the order $c_2c_1c_0$, corresponding to $q_Cq_Bq_A$ and the $CBA$ ordering of (\ref{W_orig}). In the notation of (\ref{bit_decode}), $M_B=c_1$ and $M_C=c_2$, while Alice announces $M_A=c_0$ in information rounds and $M_A=c_0\oplus1$ in retained semi-information rounds. The reconstructed bit is $s_r=M_A\oplus M_B\oplus M_C$. Successive blocks continue the same run. Finally, \texttt{Recovered} contains the reconstructed secret bits, while \texttt{Secret} gives the input sequence for comparison.

For the single-round measurements and efficiency estimates, the program evaluates all possible circuit configurations for each secret-bit value. Since these configurations are equally likely, the corresponding measured probabilities are averaged. For each secret-bit value, there are two encoding choices and 16 combinations of Alice's Hadamard-gate applications and the participants' measurement bases. Thus, a complete experiment contains 64 configurations without an attack or for a fixed entangle-and-measure (E\&M) interaction. Intercept-and-resend (I\&R) requires 128 configurations because the interception basis has two possible values. Physical choices before and after an attack are retained separately, even when they produce the same effective configuration without an attack. Each configuration is measured the same number of times, so pooling the results is equivalent to averaging the probabilities estimated for the individual configurations. The cumulative detection probabilities are evaluated separately through repeated protocol runs, as described below.

The hardware experiments were performed on \texttt{ibm\_kingston}, an IBM Quantum processor. We used 1024 shots per circuit for the efficiency experiment and 256 shots per protocol configuration for each of the two separate no-attack controls. In the intercept-and-resend experiment, we collected 256 shots per protocol configuration in each of the two interception bases, for a total of 512 shots across both bases. In the entangle-and-measure experiment, we likewise collected 512 shots per protocol configuration at each interaction angle. The separate ancilla-information experiment used 2048 shots for each input state and interaction angle. In Aer, we used 2048 shots per configuration for the single-round protocol experiments and 8192 shots per input state and interaction angle for the ancilla-information experiment. Each cumulative detection experiment consisted of 1000 independent Aer runs, each attempting to transmit a predetermined 100-bit secret. A run ended when a forbidden outcome was observed or when transmission of all 100 bits was completed, with no preset limit on the number of attempted rounds. The notebook uses no fixed random seeds. It reuses available hardware measurements and submits the corresponding circuits to the processor when a required dataset is absent.

To quantify sampling uncertainty, we construct conservative, pointwise 95\% confidence bounds by inverting Bernoulli KL/Chernoff inequalities \cite{Hoeffding1963}. Bounds on the two bit-conditioned acceptance probabilities are constructed jointly and propagated to the estimated expected number of rounds. For the ancilla experiment, mutual-information bounds are obtained by taking the minimum and maximum over joint bounds on the two conditional output probabilities. The calculations assume independent shots with stationary statistics for each configuration. These intervals describe uncertainty in the estimated probabilities and derived quantities, rather than variation in the duration of an individual protocol run.

\subsection{Analysis of efficiency}

The experiment estimates $P(R_i\in I)$, including retained semi-information rounds, as defined in Section~\ref{sec:3}. The ideal acceptance probability and expected number of rounds are given in (\ref{Prob_inf}) and (\ref{E_m}). In the experimental estimate, the probability $1/4$ of selecting an ordinary information-round configuration is fixed by the protocol choices, while measurements determine the probability that a semi-information configuration satisfies Alice's selection condition. The additional retention probability remains $1/2$.

Let $\widehat{P}_{I,s}$ denote the estimated acceptance probability conditional on the encoded bit $s\in\{0,1\}$. For a secret containing $n_0$ zeros and $n_1$ ones, independent repeated attempts with stationary acceptance probabilities give
\begin{equation}
 \widehat{E[m]}=\frac{n_0}{\widehat{P}_{I,0}}+
                  \frac{n_1}{\widehat{P}_{I,1}},
 \qquad n_0+n_1=n.
 \label{impl_efficiency}
\end{equation}
For the plotted comparison, we take $n_0=n_1=n/2$, giving $\widehat{E[m]}=(n/2)(1/\widehat{P}_{I,0}+1/\widehat{P}_{I,1})$. Results are shown for $n=10,20,40,80$.

\begin{figure}[htbp]
 \centering
 \includegraphics[width=0.8\linewidth]{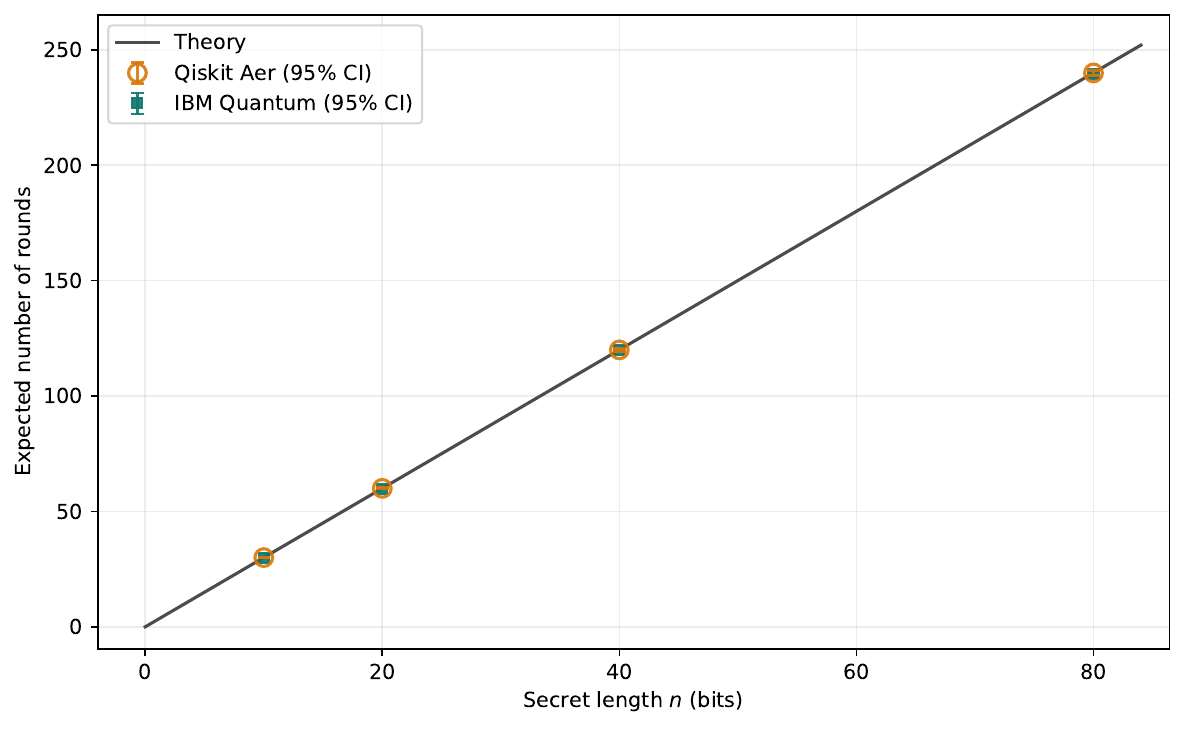}
 \caption{Estimated expected number of protocol rounds required to transmit an $n$-bit secret. The solid line represents $E[m]=3n$. Error bars show pointwise 95\% confidence bounds.}
 \label{fig:implementation_efficiency}
\end{figure}

Figure~\ref{fig:implementation_efficiency} compares the experimental estimates with the theoretical expectation (\ref{E_m}). The points within each experimental series are calculated from the same estimated acceptance probabilities using (\ref{impl_efficiency}), which gives a linear dependence on $n$. The hardware measurements give $\widehat{E[m]}=2.9945n$, corresponding to approximately $2.9945$ protocol rounds per transmitted secret bit. The conservative 95\% confidence interval for this mean number of rounds is $[2.9776,3.0120]$, which contains the theoretical value of three.

These estimates describe the number of rounds required for secret transmission without conditioning on acceptance of the security tests. Retaining a round for transmission does not ensure correct reconstruction of the encoded bit on hardware. Applying the decoding rule (\ref{bit_decode}) to the same no-attack measurements gives an estimated error fraction of approximately $4.30\%$ in the reconstructed secret, corresponding to its quantum bit error rate (QBER). This value is obtained by averaging the error fractions calculated separately for encoded bits 0 and 1, with semi-information retention included. It characterizes reconstruction before any subsequent error correction or conditioning on a successful security test.
\subsection{Intercept-and-resend attack}

Bob's interception of Charlie's qubit is implemented by a projective measurement in a uniformly chosen Z or X basis. For interception in the X basis, H is applied before the interception measurement and again afterwards. The resulting state of the same circuit qubit is the eigenstate that Bob would resend, implementing the ideal measure-and-resend channel. The interception result is recorded separately and excluded from the three-bit protocol outcome.

\begin{figure}[htbp]
 \centering
 \includegraphics[width=0.8\linewidth]{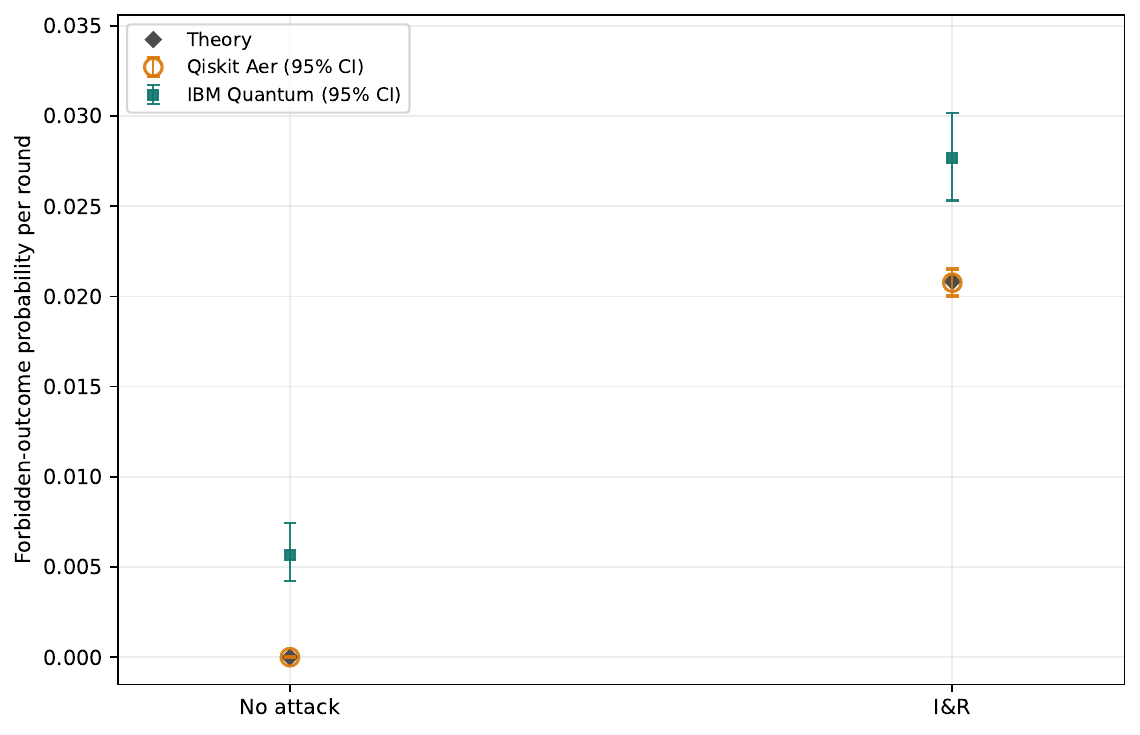}
 \caption{Forbidden-outcome probability per protocol round without an attack and under I\&R. Error bars show pointwise 95\% confidence bounds.}
 \label{fig:implementation_ir}
\end{figure}

For each test configuration, the program compares the measurement outcome with the forbidden set for the corresponding encoded state. Averaging over the protocol choices and the interception basis gives the probability $P_{\mathrm{det}}=1/48$ derived in (\ref{prob_det_gen}). Figure~\ref{fig:implementation_ir} compares this ideal prediction with the simulator and hardware frequencies.

The Aer results are close to the ideal prediction. On \texttt{ibm\_kingston}, the estimated forbidden-outcome probability increases from $0.00568$ without an attack to $0.02768$ under intercept-and-resend. The hardware attack frequency exceeds the ideal value $1/48\approx0.02083$, while the no-attack control exhibits a nonzero background. Thus, forbidden outcomes remain sensitive to the intervention, but an isolated hardware outcome is insufficient to distinguish an attack from the clean-device background. A calibrated statistical criterion is introduced in Subsection~\ref{sec:4_statistical}.

We next examine the cumulative detection probability derived in Section~\ref{sec:3}: the probability of observing a forbidden outcome before transmission of an $n$-bit secret is completed. Each information round and each retained semi-information round is used to transmit one bit of the predetermined secret. The number of attempted protocol rounds is random.

The simulator evaluates this probability using 1000 independent protocol runs per attack setting. Before each run, a random secret of the largest considered length is chosen. The current bit is repeatedly encoded in freshly prepared states until an information round or a retained semi-information round is obtained. A forbidden test outcome aborts the run; otherwise, it continues until transmission of the entire secret is completed. There is no fixed cutoff on the number of attempted rounds. Circuit executions are grouped to reduce simulator overhead, with new batches acquired when necessary. Every consumed measurement outcome comes from a distinct fresh Aer shot.

For each protocol run $r$, let $L_r$ denote the number of information rounds and retained semi-information rounds obtained before termination. Each such round is used to transmit one secret bit. Since a run stops only upon a forbidden outcome or completion of the maximum considered secret length, the condition $L_r<n$ means that detection occurred before transmission of the first $n$ secret bits was completed. Thus,
\begin{equation}
 \widehat P_{\mathrm{det}}^{(n)}
 =\frac{1}{N_{\mathrm{run}}}
   \sum_{r=1}^{N_{\mathrm{run}}}\mathbf{1}\{L_r<n\},
 \qquad N_{\mathrm{run}}=1000.
 \label{impl_empirical_cumulative}
\end{equation}
Here, $\mathbf{1}\{L_r<n\}$ is the indicator function, equal to one if $L_r<n$ and zero otherwise. Thus, the sum counts only runs aborted before transmission of the first $n$ secret bits was completed. The plotted estimates therefore measure the frequencies of aborted transmissions directly. Their error bars are pointwise 95\% Bernoulli bounds across independent protocol runs. The same protocol runs are used for different secret lengths, so the plotted points are correlated and the bounds do not provide simultaneous coverage of the whole curve. At $n=0$, no transmission is required and the detection probability is exactly zero.

\begin{figure}[htbp]
 \centering
 \includegraphics[width=0.8\linewidth]{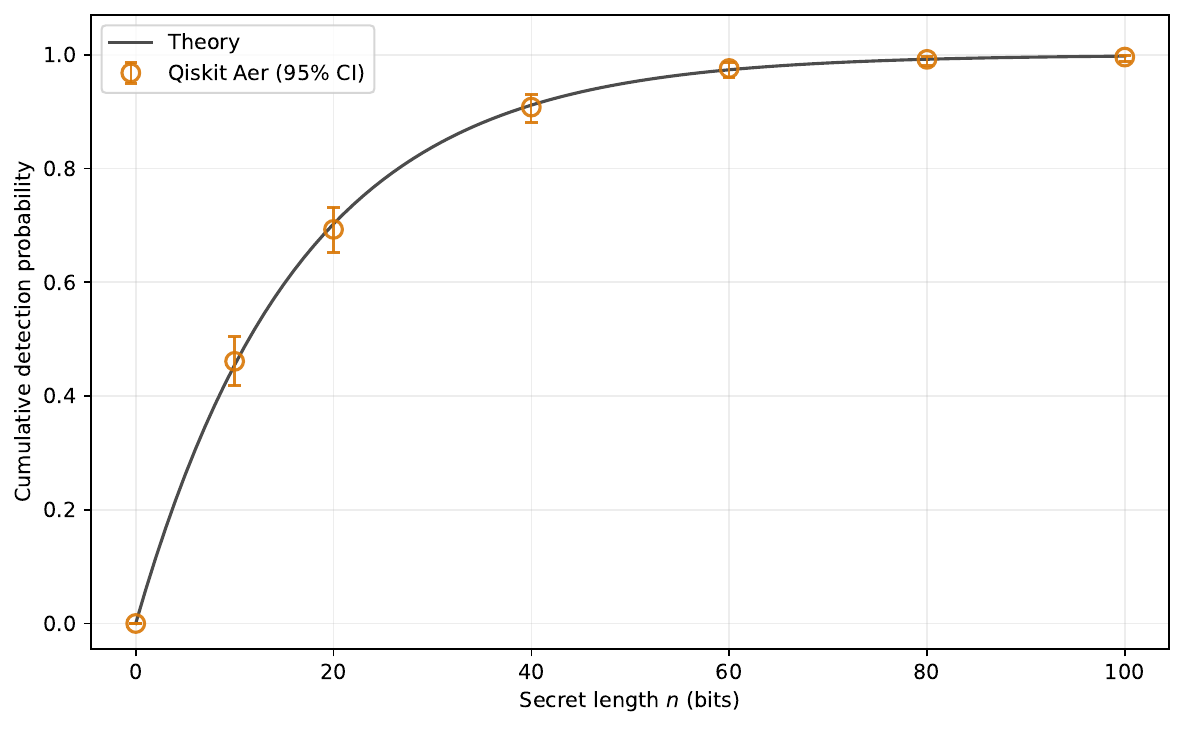}
 \caption{Cumulative I\&R detection probability before transmission of an $n$-bit secret is completed. The solid line gives (\ref{prob_det_ir_cumulative}); open circles show abort frequencies from 1000 independent protocol runs on Aer. Error bars show pointwise 95\% confidence bounds.}
 \label{fig:implementation_ir_cumulative}
\end{figure}

Figure~\ref{fig:implementation_ir_cumulative} compares the theoretical probability with the simulator frequencies for $n=0,10,20,40,60,80,100$. At $n=100$, (\ref{prob_det_ir_cumulative}) gives a probability of approximately $0.998$. Here, $n$ denotes the secret length, with both information rounds and retained semi-information rounds contributing to its transmission.

\subsection{Entangle-and-measure attack}

The numerical experiment implements a particular member of the interaction class considered in Section~\ref{sec:3}. An additional qubit represents Bob's ancillary system $E$ and is initialized in $\ket{E_0}_E=\ket{0}_E$. A controlled $R_y(2\theta)$ gate, with Charlie's qubit as control, implements
\begin{equation}
\begin{aligned}
 U_{CE}(\theta)\ket{0}_C\ket{E_0}_E
 &= \ket{0}_C\ket{e_0}_E = \ket{0}_C\ket{0}_E,\\
 U_{CE}(\theta)\ket{1}_C\ket{E_0}_E
 &= \ket{1}_C\ket{e_3}_E
  = \ket{1}_C\bigl(\cos\theta\ket{0}_E+\sin\theta\ket{1}_E\bigr).
\end{aligned}
\label{impl_em_interaction}
\end{equation}
Here $\ket{e_1}=\ket{e_2}=0$ and $\braket{e_0}{e_3}=\cos\theta$. Substitution into (\ref{prob_det_em_gen}) therefore gives
\begin{equation}
 P_{\mathrm{det}}(\theta)=\frac{1-\cos\theta}{48}.
 \label{impl_em_detection}
\end{equation}
The interaction is applied after Alice's Hadamard operations and before the participants' measurements. Experiments use $\theta\in\{0,\pi/4,\pi/3,\pi/2,3\pi/4,\pi\}$.

\begin{figure}[htbp]
 \centering
 \includegraphics[width=0.8\linewidth]{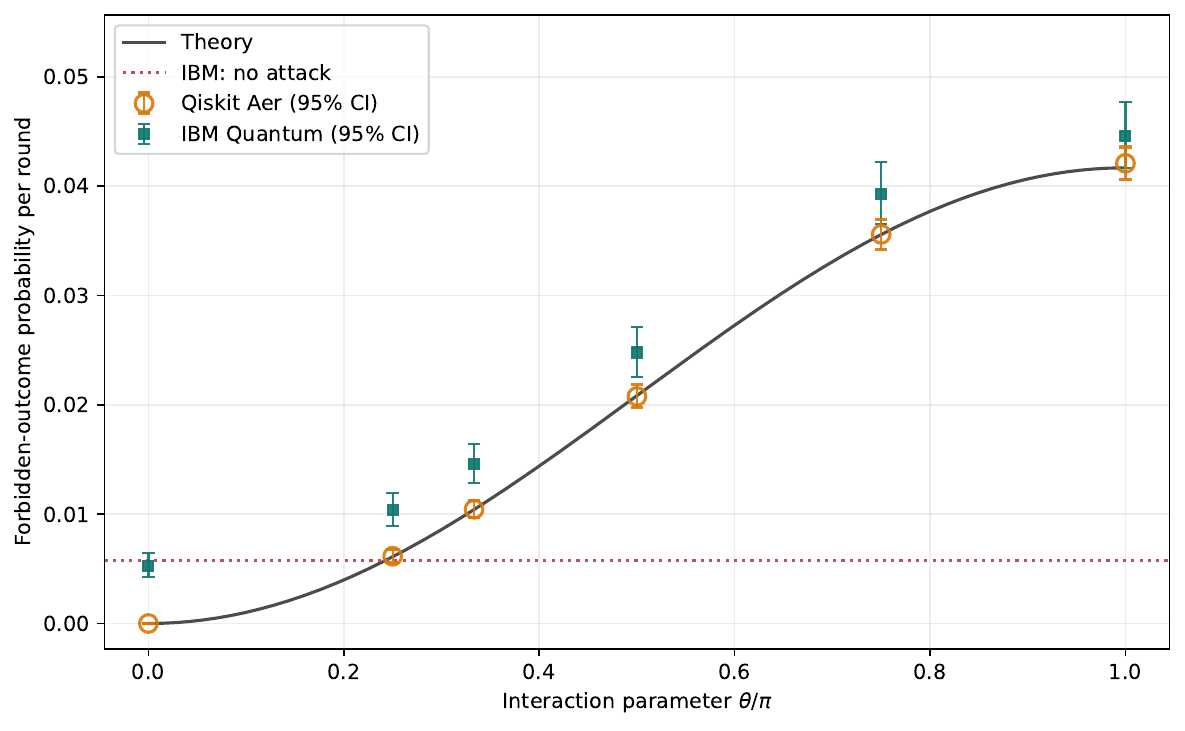}
 \caption{Forbidden-outcome probability per protocol round under E\&M as a function of $\theta$. The dotted line gives the separate hardware no-attack control. Error bars show pointwise 95\% confidence bounds.}
 \label{fig:implementation_em_detection}
\end{figure}

Figure~\ref{fig:implementation_em_detection} shows that the simulator reproduces (\ref{impl_em_detection}). The hardware results exhibit an overall increase in the forbidden-outcome probability as $\theta$ increases, reaching $0.04459$ at $\theta=\pi$, compared with the ideal value $1/24\approx0.04167$. The separate hardware control has a forbidden-outcome probability of $0.00574$. Agreement with the ideal curve is not uniform: at $\theta=\pi/4$, the measured probability is $0.01038$, with confidence interval $[0.00893, 0.01196]$, above the ideal prediction of approximately $0.00610$.

The cumulative probability is evaluated using the same stopping rule as for intercept-and-resend. Since $\|\ket{e_0}-\ket{e_3}\|^2=2(1-\cos\theta)$, substitution into (\ref{prob_det_em_cumulative}) gives
\begin{equation}
 P_{\mathrm{det}}^{(n)}(\theta)
 =1-\left(\frac{16}{17-\cos\theta}\right)^n.
 \label{impl_em_cumulative}
\end{equation}
The same simulation procedure is repeated at $\theta=\pi/4,\pi/3,\pi/2,3\pi/4,\pi$, with 1000 independent protocol runs for each parameter. The value $\theta=0$ gives no interaction and an exactly zero ideal detection probability, so its cumulative curve is omitted.

\begin{figure}[htbp]
 \centering
 \includegraphics[width=0.8\linewidth]{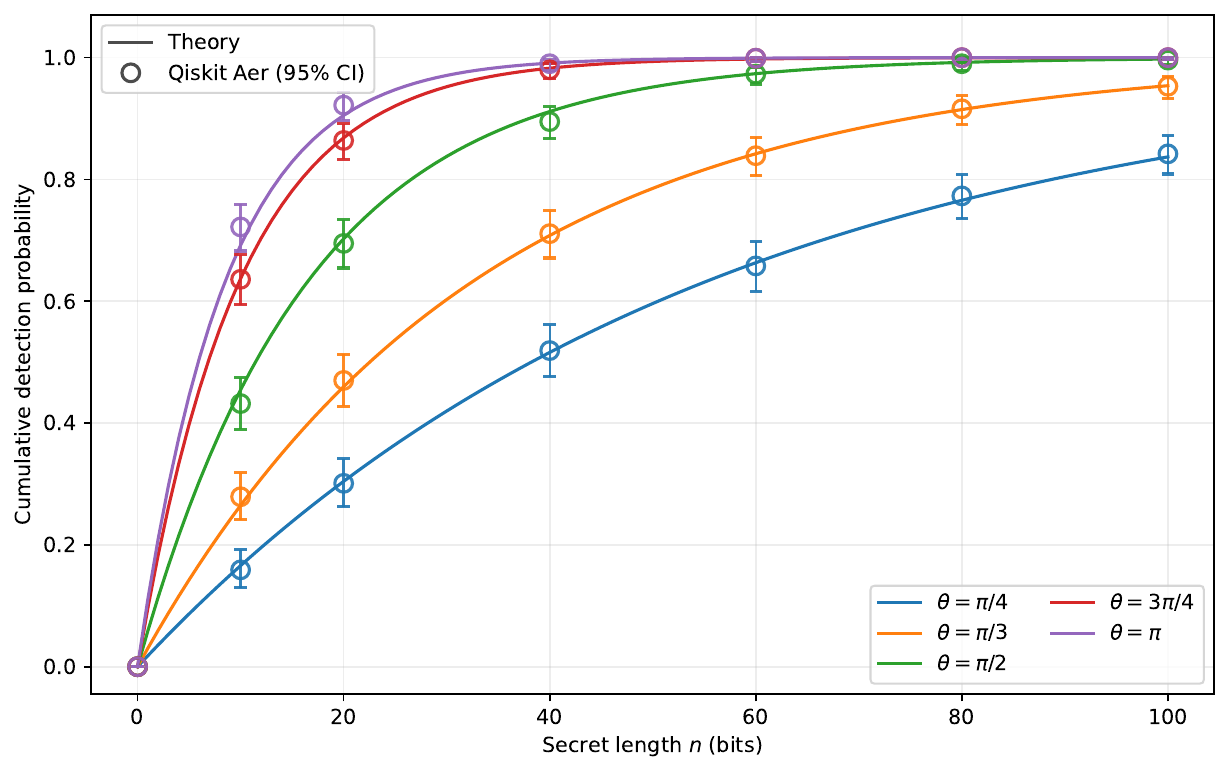}
 \caption{Cumulative E\&M detection probability before transmission of an $n$-bit secret is completed for different values of $\theta$. Solid lines give (\ref{impl_em_cumulative}); open circles show abort frequencies from 1000 independent protocol runs on Aer per interaction parameter. Colors identify $\theta$, and error bars show pointwise 95\% confidence bounds.}
 \label{fig:implementation_em_cumulative}
\end{figure}

Figure~\ref{fig:implementation_em_cumulative} shows that interactions producing greater single-round disturbance are more likely to be detected before transmission of a secret of a given length is completed. At $\theta=\pi/2$, the ideal per-round detection probability equals the intercept-and-resend value, and the theoretical cumulative curves coincide. Even the weaker interaction at $\theta=\pi/4$ gives a cumulative probability of approximately $0.837$ for a 100-bit secret. These curves evaluate the ideal rule in which a single forbidden outcome detects the intervention. On a noisy processor, clean-device errors can also cause an abort. The hardware experiments are therefore assessed separately using the calibrated criterion in Subsection~\ref{sec:4_statistical}.

A separate experiment characterizes the information about the prepared input that can be obtained by measuring the ancilla. The program prepares the inputs $\ket{0}_C$ and $\ket{1}_C$ separately and applies (\ref{impl_em_interaction}) in each case. The classical variable $J$ labels the prepared input: $J=0$ corresponds to $\ket{0}_C$ and $J=1$ to $\ket{1}_C$. Both inputs are assigned equal prior probabilities. The ancilla is then measured in the Z basis after applying $R_y(\pi/2-\theta)$, and its binary outcome is denoted by $Y$.

For each input $j$, the circuit is executed repeatedly. Let $N_{j,y}$ be the number of measurements yielding the ancilla outcome $y$, and let $N_j=\sum_y N_{j,y}$ be the total number of measurements for that input. The frequency $N_{j,y}/N_j$ estimates the conditional probability of observing $Y=y$ given $J=j$. The empirical joint distribution is therefore constructed as
\begin{equation}
 \widehat p(j,y)=\frac12\frac{N_{j,y}}{N_j}.
\end{equation}
The factor $1/2$ accounts for the equal prior probabilities of the two inputs. The four values of $\widehat p(j,y)$ are then used to calculate the measured mutual information $\widehat I(J;Y)$, which quantifies how much the ancilla outcome reveals about the prepared input. This input pair is used to characterize the chosen interaction and does not restrict the general interaction model of Section~\ref{sec:3}.

For the ideal ancillary states, $D=|\sin\theta|$, and the specified measurement achieves
\begin{equation}
 I_{\mathrm{acc}}(\theta)
 =1-h_2\!\left(\frac{1-|\sin\theta|}{2}\right),
 \label{impl_em_information}
\end{equation}
where $h_2$ is the binary entropy defined in Section~\ref{sec:3}. On hardware, the experiment measures the mutual information achieved by this particular measurement. Its optimality for the noisy ancillary states is not assumed. The quantity concerns distinguishability of the conditional ancilla states, rather than the total information available to Bob about the encoded secret, which can also involve his own qubit and classical announcements.

\begin{figure}[htbp]
 \centering
 \includegraphics[width=0.8\linewidth]{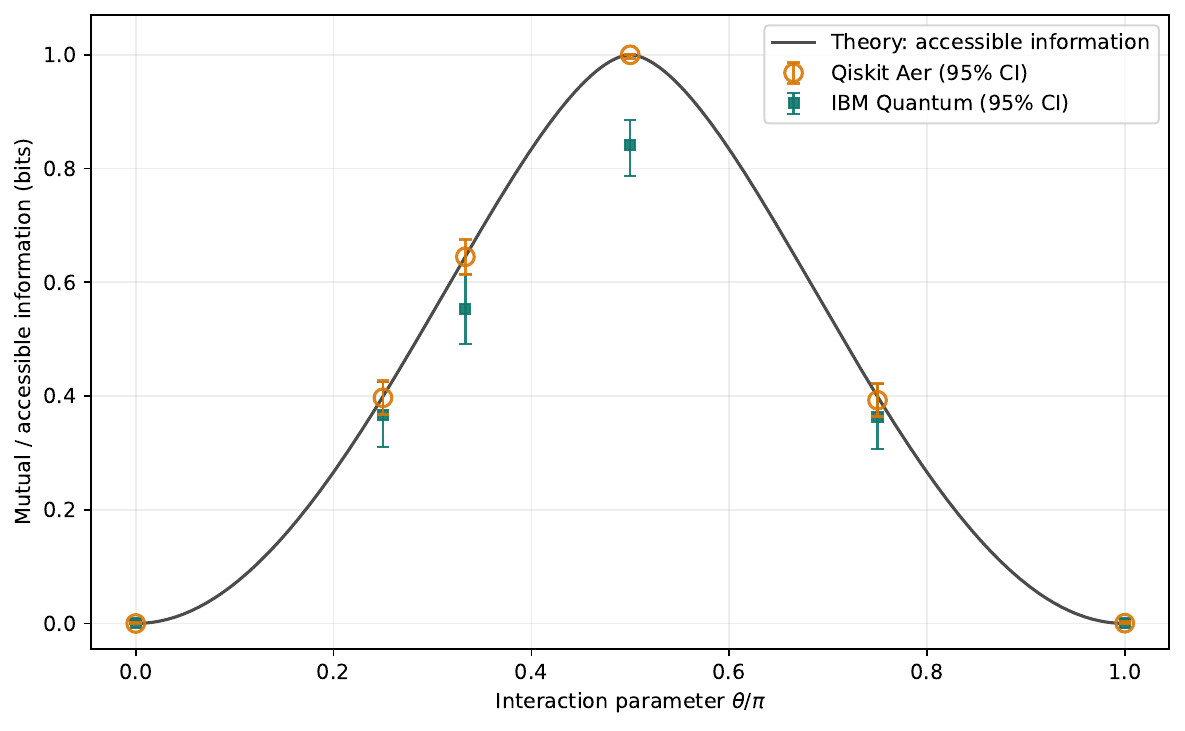}
 \caption{Ideal accessible information and measured mutual information between the prepared input state and the ancilla measurement outcome under E\&M. Error bars show pointwise 95\% confidence bounds.}
 \label{fig:implementation_em_information}
\end{figure}

Figure~\ref{fig:implementation_em_information} shows the predicted maximum at $\theta=\pi/2$ and the reduction towards both endpoints. The hardware mutual information at $\theta=\pi/2$ is approximately $0.842$ bits, with confidence interval $[0.788,0.886]$, below the ideal value of one bit. At $\theta=0$ and $\theta=\pi$, the measured information is close to zero.

\begin{figure}[htbp]
 \centering
 \includegraphics[width=0.8\linewidth]{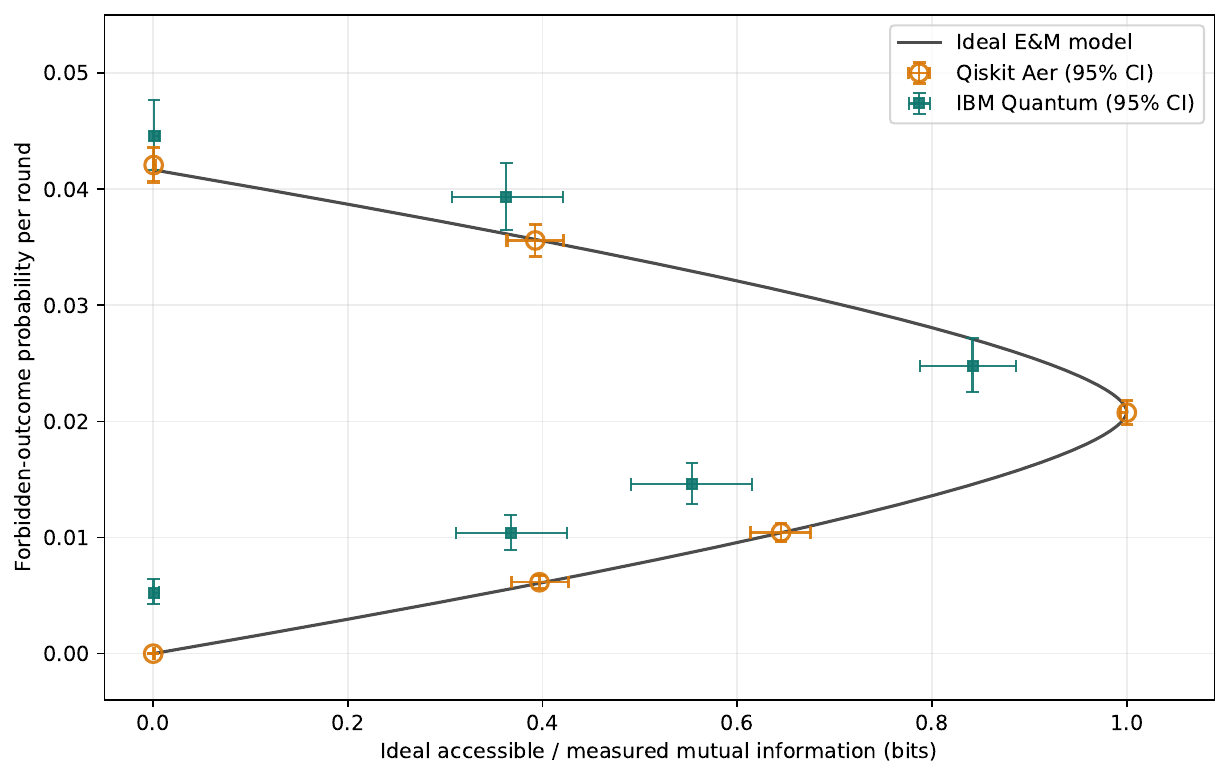}
 \caption{Information and disturbance at matching values of $\theta$. The ideal curve is parameterized by (\ref{impl_em_detection}) and (\ref{impl_em_information}); experimental points combine the corresponding measurements. Horizontal and vertical error bars show marginal 95\% confidence bounds.}
 \label{fig:implementation_information_disturbance}
\end{figure}

The two branches in Figure~\ref{fig:implementation_information_disturbance} arise because $\theta$ and $\pi-\theta$ give the same ideal accessible information but different disturbance probabilities. The lower branch, $0\le\theta\le\pi/2$, attains the information--disturbance bound derived in Section~\ref{sec:3}. At $\theta=\pi$, $\ket{e_3}=-\ket{e_0}$, so the ancillary density operators coincide and provide no distinguishing information. Nevertheless, the interaction induces a phase change on Charlie's qubit and yields nonzero disturbance. Consequently, absence of ancilla information does not imply absence of disturbance.

\subsection{Statistical attack detection}\label{sec:4_statistical}

The ideal protocol aborts when any forbidden test outcome is observed. The hardware controls show that such outcomes can also arise from device noise, so the presence of a forbidden outcome alone is insufficient to identify an attack. We therefore examine whether the observed count exceeds what can reasonably be attributed to the calibrated no-attack background. The following criterion first estimates this background and then uses an upper confidence bound to select a detection threshold. It is evaluated separately from the ideal accept-or-abort demonstration.

Let $m_T=|T|$ be the number of test rounds and let $K$ count their forbidden outcomes. The total number of protocol rounds retains the notation $m$ from Section~\ref{sec:3}. Each test-round outcome is checked against the forbidden set for its configuration before the forbidden outcomes are counted across configurations. Since $P(R_i\in T)=1/2$, the ideal probability conditional on membership in the complete test set is
\begin{equation}
 q_{\mathrm{ideal}}
 =P(\text{forbidden outcome}\mid R_i\in T)
 =2P_{\mathrm{det}}
 =\begin{cases}
  1/24, & \text{I\&R},\\
  (1-\cos\theta)/24, & \text{E\&M}.
 \end{cases}
 \label{impl_pooled_test_probability}
\end{equation}
This conditioning includes both test-round types. It therefore differs from $P_{\mathrm{det}\mid\mathrm{test}}$ in (\ref{prob_det}) and (\ref{prob_det_em}), which is evaluated for the specific detecting configurations analyzed in Section~\ref{sec:3}.

For the clean hardware model, let $q_s$ denote the probability of a forbidden result conditional on a test round and encoded bit $s$. The null hypothesis is that subsequent test rounds follow these clean probabilities, with the same configuration sampling and stationary device statistics.  For each attack family, we randomly split the no-attack shots within each configuration into two disjoint subsets of equal size, one for calibration and one for holdout. The calibration subset contains $N_{\mathrm{cal},s}=2048$ test shots for each secret-bit value in both experiments. The holdout subset contains the same number of test shots and is used to estimate the no-attack probabilities for the plotted false-alarm projections.

Since the background probability is estimated from a finite sample, using its observed frequency alone could underestimate clean-device fluctuations. We therefore construct an upper confidence bound and use it when selecting the detection threshold.

Write $K_{\mathrm{cal},s}$ for the calibration forbidden count and $\widehat q_{\mathrm{cal},s}=K_{\mathrm{cal},s}/N_{\mathrm{cal},s}$. Define the Bernoulli relative entropy, using natural logarithms, by
\begin{equation}
 d(x\Vert y)=x\ln\frac{x}{y}+(1-x)\ln\frac{1-x}{1-y},
 \label{impl_bernoulli_kl}
\end{equation}
with the usual continuous endpoint conventions. For independent Bernoulli calibration shots, including unequal probabilities across configurations, the lower-tail Chernoff bound gives
\begin{equation}
 \Pr(\widehat q_{\mathrm{cal},s}\le x)
 \le \exp[-N_{\mathrm{cal},s}d(x\Vert q_s)],
 \qquad 0\le x<q_s.
 \label{impl_calibration_tail}
\end{equation}
Since the test configurations are equally likely and measured equally often, the average calibration probability is $q_s$.

To obtain joint calibration coverage of at least $1-\delta$, we allocate a failure probability of $\delta/2$ to each bit value. The corresponding upper confidence bound $u_s$ is obtained, when $0<K_{\mathrm{cal},s}<N_{\mathrm{cal},s}$, by solving
\begin{equation}
 N_{\mathrm{cal},s}
 d(\widehat q_{\mathrm{cal},s}\Vert u_s)
 =\ln\frac{2}{\delta},
 \qquad \widehat q_{\mathrm{cal},s}<u_s<1.
 \label{impl_calibration_upper}
\end{equation}
If $K_{\mathrm{cal},s}=0$, then $u_s=1-(\delta/2)^{1/N_{\mathrm{cal},s}}$; if $K_{\mathrm{cal},s}=N_{\mathrm{cal},s}$, then $u_s=1$. Inverting (\ref{impl_calibration_tail}) gives $\Pr(q_s>u_s)\le\delta/2$. A union bound over the two bit values therefore yields
\begin{equation}
 \Pr(\mathcal G)\ge1-\delta,
 \qquad
 \mathcal G=\{q_0\le u_0,\ q_1\le u_1\}.
 \label{impl_calibration_coverage}
\end{equation}
The experiment uses $\delta=0.05$ and sets $u=\max(u_0,u_1)$, covering both encoded bit values without assuming that their clean rates are identical.

Once the background bound $u$ has been obtained, we choose a critical count so that reaching or exceeding it is sufficiently unlikely under the clean model. For $B\sim\operatorname{Bin}(m_T,u)$, define the upper tail
\begin{equation}
 \mathcal B_{m_T}(k;u)
 =\Pr(B\ge k)
 =\sum_{j=k}^{m_T}\binom{m_T}{j}u^j(1-u)^{m_T-j}.
\end{equation}
The critical count at conditional significance level $\alpha$ is
\begin{equation}
 k_{\mathrm{crit}}(m_T,\alpha)
 =\min\{k\in\{1,\ldots,m_T+1\}:
                 \mathcal B_{m_T}(k;u)\le\alpha\}.
 \label{impl_threshold}
\end{equation}
The empty tail at $k=m_T+1$ is zero. The decision rule declares statistically significant excess disturbance if and only if
\begin{equation}
 K\ge k_{\mathrm{crit}}(m_T,\alpha).
 \label{impl_decision}
\end{equation}

To justify this rule, consider a calibration result satisfying $\mathcal G$. After averaging over independently sampled test configurations, a clean test round with encoded bit $s$ yields a forbidden outcome with probability $q_s\le u$. Under independent subsequent rounds, the number of forbidden outcomes is stochastically dominated by $\operatorname{Bin}(m_T,u)$, even for unequal bit-conditioned rates. Consequently, for every such calibration result, the clean probability of crossing the threshold is at most $\alpha$. Accounting also for calibration failure gives
\begin{equation}
 \Pr_0(\text{declare excess disturbance})
 \le \alpha(1-\delta)+\delta
 =\alpha+(1-\alpha)\delta.
 \label{impl_overall_false_alarm}
\end{equation}
Here $\Pr_0$ refers to the stationary clean model. Thus, $\alpha$ is a conditional testing level, while $\delta$ controls calibration coverage. With $\delta=0.05$, the conservative unconditional bounds are $0.0595$ for $\alpha=0.01$ and $0.0975$ for $\alpha=0.05$; the plotted false-alarm estimates are not these worst-case bounds.

The guarantee also applies to a single decision at the end of transmission, when the number of test rounds is random. This requires a secret sequence fixed before transmission, independent attempts with stationary statistics for each encoded bit and collection that does not stop in response to forbidden outcomes. Conditioning on the sequence of test, information and discarded rounds fixes the number of test rounds and the secret-bit value encoded in each of them, so the same binomial bound applies. Repeated outcome-dependent monitoring requires a separate analysis.

For the numerical power study, let $\widehat q_{\mathrm{atk}}$ and $\widehat q_{\mathrm{hold}}$ denote the test-round forbidden-outcome probabilities estimated from the attack and clean holdout measurements, respectively. Each estimate is the mean of the corresponding probabilities calculated separately for encoded bits 0 and 1. The plotted detection and false-alarm probabilities are
\begin{align}
 \widehat\pi_{\mathrm{atk}}(m_T,\alpha)
 &=\mathcal B_{m_T}\bigl(k_{\mathrm{crit}};
                              \widehat q_{\mathrm{atk}}\bigr),\\
 \widehat\pi_0(m_T,\alpha)
 &=\mathcal B_{m_T}\bigl(k_{\mathrm{crit}};
                              \widehat q_{\mathrm{hold}}\bigr).
 \label{impl_power}
\end{align}
These probabilities are calculated using a binomial model with the measured forbidden-outcome frequencies. They are not estimated by counting threshold crossings across repeated complete protocol runs. They describe a model in which successive test draws are independent and the two encoded bit values occur with equal probability. They are therefore not exact power calculations for every fixed secret sequence. The calibration partition is randomly generated in each notebook run, so exact threshold curves can vary between runs. The null bound uses $u$, not the holdout point estimate.

\begin{figure}[htbp]
 \centering
 \includegraphics[width=0.8\linewidth]{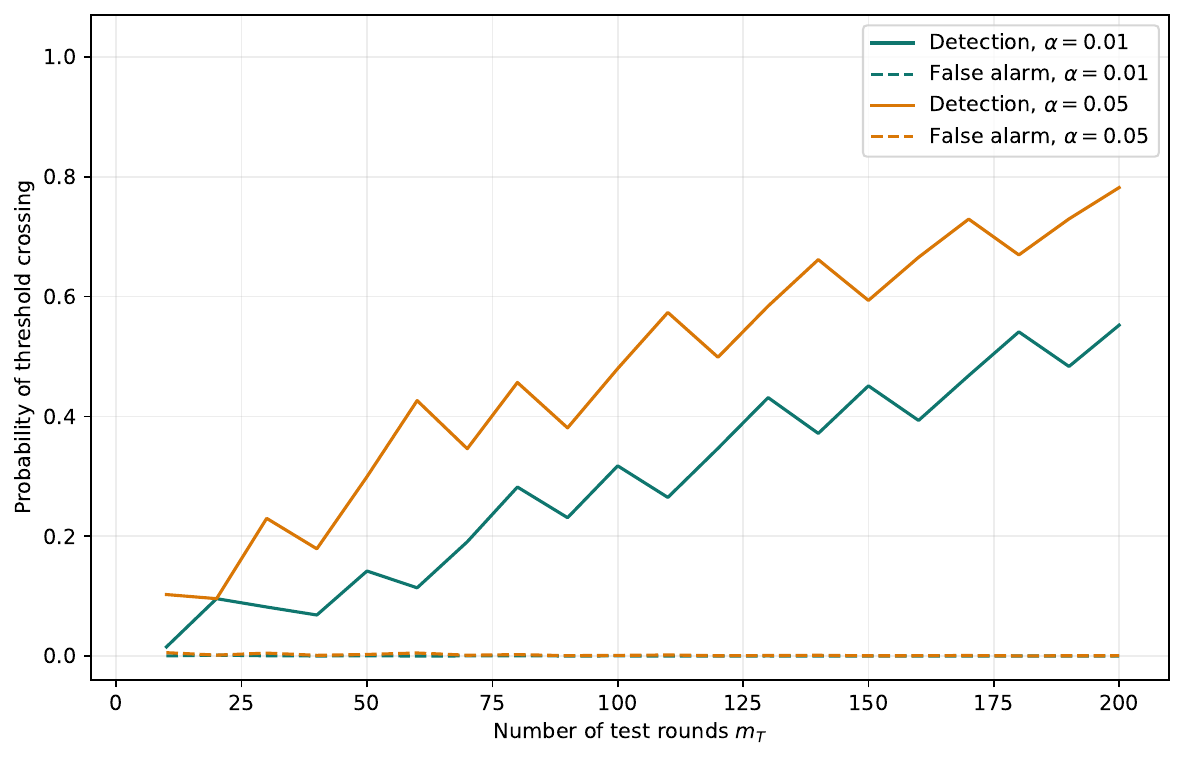}
 \caption{Projected I\&R detection and clean false-alarm probabilities versus the number of test rounds $m_T$, at conditional levels $\alpha=0.01$ and $0.05$.}
 \label{fig:implementation_ir_threshold}
\end{figure}

Figure~\ref{fig:implementation_ir_threshold} evaluates $m_T=10,20,\ldots,200$. For the displayed calibration split, the projected intercept-and-resend detection probability at 200 test rounds is approximately $0.5523$ for $\alpha=0.01$ and $0.7819$ for $\alpha=0.05$. The projected clean false-alarm probabilities remain below the corresponding conditional levels. The local decreases in power occur when the integer critical count increases; monotonic growth at every successive test count is not required for this non-randomized criterion.

\begin{figure}[H]
 \centering
 \includegraphics[width=\linewidth]{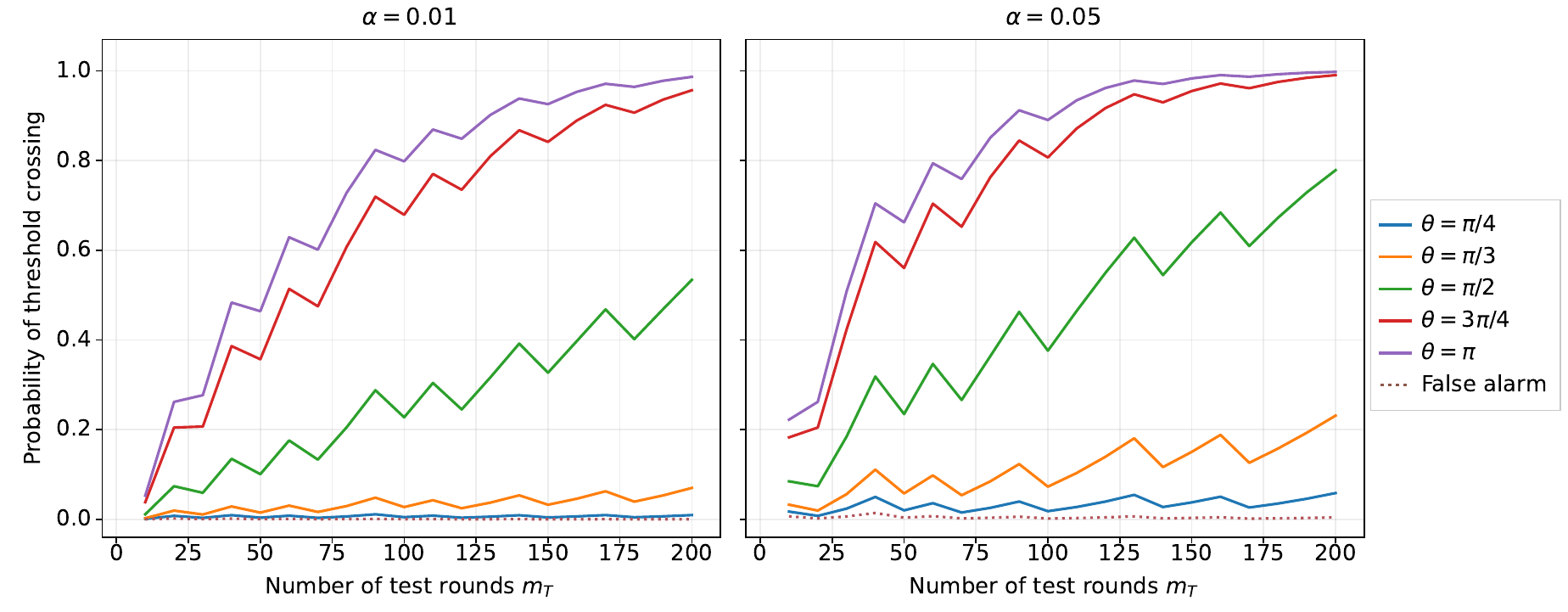}
 \caption{Projected E\&M detection probabilities for different interaction parameters versus the number of test rounds $m_T$. The dotted curves show the clean holdout projections. The two panels use conditional levels $\alpha=0.01$ and $0.05$.}
 \label{fig:implementation_em_threshold}
\end{figure}

Figure~\ref{fig:implementation_em_threshold} shows high projected detection probabilities at the larger interaction angles $\theta=3\pi/4$ and $\theta=\pi$, while sensitivity at $\theta=\pi/4$ and $\theta=\pi/3$ remains low over the considered range. At $\theta=\pi/2$ and $m_T=200$, the projected detection probability is approximately $0.5340$ for $\alpha=0.01$ and $0.7786$ for $\alpha=0.05$. Thus, nonzero information and nonzero ideal disturbance do not imply reliable detection from a small finite hardware sample using this criterion.

A threshold crossing identifies disagreement with the calibrated clean model. It does not determine the physical cause of that disagreement. The conclusions concern the implemented attacks and the stated statistical assumptions; they do not establish general security of a noisy protocol against arbitrary strategies.

\section{Comparison with related secret-sharing protocols}\label{sec:5}

Table~\ref{tab:qss_comparison} compares the proposed scheme with selected protocols for sharing classical secrets. The first three entries use W states, while the remaining literature entries provide examples based on GHZ or Bell states. The comparison distinguishes how the secret is obtained, what operations the receivers perform and how the protocol checks for eavesdropping.

\begingroup
\footnotesize
\setlength{\tabcolsep}{1.4pt}
\renewcommand{\arraystretch}{1.12}

\begin{longtable}{cccccc}
\toprule
Protocol &
Resource &
Secret type &
\makecell{Receiver operations\\and return} &
Security check &
\makecell{Qubit\\efficiency} \\
\midrule
\endfirsthead

\toprule
Protocol &
Resource &
Secret type &
\makecell{Receiver operations\\and return} &
Security check &
\makecell{Qubit\\efficiency} \\
\midrule
\endhead

\makecell{Liu et al.\\\cite{LiuTsaiHwang2012W}} &
\makecell{Symmetric W\\states, decoys} &
Predetermined &
\makecell{$Z$ for recovery\\No return} &
Decoy tests &
$\dfrac{n}{3n+D}$ \\
\addlinespace

\makecell{Tsai et al.\\\cite{TsaiYangLee2019W}} &
\makecell{Three-qubit\\W states} &
\makecell{Measurement-\\generated} &
\makecell{$Z$ measure-and-\\resend or reflection} &
\makecell{Bell-basis\\measurements and\\sampled key bits} &
$1/8$ \\
\addlinespace

\makecell{Xing et al.\\\cite{Xing2025W}} &
\makecell{Three-qubit\\W states, decoys} &
Predetermined &
\makecell{$Z$ for recovery\\Decoys shuffled\\and returned} &
\makecell{Decoy tests and\\sampled bits} &
$\dfrac{n}{3n+2l}$ \\
\addlinespace

\makecell{Hillery et al.\\\cite{Hillery1999}} &
\makecell{Three-qubit\\GHZ states} &
\makecell{Measurement-\\generated} &
\makecell{$X$ or $Y$\\No return} &
\makecell{Comparison of\\sampled outcomes} &
$1/6$\textsuperscript{a} \\
\addlinespace

\makecell{Xie et al.\\\cite{Xie2015Specific}} &
\makecell{Three-qubit\\GHZ-type states} &
Predetermined &
\makecell{$Z$ measure-and-\\resend or reflection} &
\makecell{Joint measurements\\by Alice} &
$\dfrac{1-(3/4)^N}{3N}$\textsuperscript{b} \\
\addlinespace

\makecell{Tian et al.\\\cite{Tian2021Specific}} &
\makecell{Bell pairs,\\decoys} &
Predetermined\textsuperscript{c} &
\makecell{$Z$ for recovery\\Only decoys returned} &
\makecell{Decoy tests and\\sampled bits} &
$\approx 0.5$ \\
\midrule

\makecell{Proposed\\protocol} &
\makecell{Three-qubit\\W states} &
Predetermined &
\makecell{$Z$ or $X$\\No return} &
\makecell{Forbidden outcomes\\in test rounds} &
$1/9$ (\ref{eff}) \\
\bottomrule

\caption{Comparison of selected protocols for sharing classical secrets.}
\label{tab:qss_comparison}
\end{longtable}

\noindent
\begin{minipage}{\linewidth}
\footnotesize
\textsuperscript{a}~Calculated by counting all three qubits per GHZ state and retaining half of the rounds, before disclosing bits for security testing.

\textsuperscript{b}~Calculated including block restarts in the ideal no-attack model, as explained below.

\textsuperscript{c}~As stated by Tian et al., although their first step specifies random selection of Bell states.

\end{minipage}

\endgroup
\noindent
\newline
The table compares the original versions of the protocols. Security vulnerabilities in the schemes of Hillery et al., Xie et al. and Tian et al. were identified in subsequent studies \cite{Qin2007HBB,YinFu2016Xie,He2024}.

Among the W-state schemes, those of Liu et al.~\cite{LiuTsaiHwang2012W} and Xing et al.~\cite{Xing2025W} are the closest in purpose because both share a predetermined classical secret. Both introduce separate decoy photons for security checks. In Liu et al., the receivers report their decoy measurement outcomes, whereas in Xing et al. they return reflected or newly prepared decoy photons to Alice. In the proposed protocol, Alice's Hadamard-gate choices and the receivers' measurement bases determine which W-state rounds are used for testing. The test uses outcomes forbidden in the ideal protocol, and the quantum transmission proceeds from Alice to the receivers.

Tsai et al.~\cite{TsaiYangLee2019W} and Xing et al.~\cite{Xing2025W} allow the receivers to perform all their measurements in the Z basis. Both schemes use reflection and the preparation of Z-basis states for resending. Xing et al. also require the receivers to randomly permute the decoy photons. In the proposed protocol, Bob and Charlie measure in either the Z or the X basis and do not return qubits. The GHZ-based schemes illustrate two distinct approaches. Hillery et al.~\cite{Hillery1999} generate shared bits from local measurement outcomes, whereas Xie et al.~\cite{Xie2015Specific} encode a predetermined bit and use joint measurements on returned qubits for the security check.

The efficiency in (\ref{eff}) counts all qubits prepared for the proposed protocol, including Alice's qubits and those prepared in test rounds and discarded semi-information rounds. For Liu et al., Hillery et al. and Xie et al., the efficiencies are calculated here by counting qubits initially prepared by the dealer; additional qubit preparations by the receivers are not included. The efficiencies for Tsai et al., Xing et al. and Tian et al. are reproduced as reported in the original sources.

For the three-party version of Liu et al., $n$ is the secret length and $D$ is the total number of decoy photons, giving $n/(3n+D)$. For Xing et al., $n$ is the number of initially encoded secret bits and $l$ is the number of decoy photons per transmitted sequence. In Xie et al., each block of $N$ three-qubit states encodes one predetermined bit. The block is repeated if there is no position where both receivers measure. Since a block contains such a position with probability $1-(3/4)^N$, accounting for these restarts gives the efficiency shown in Table~\ref{tab:qss_comparison} in the ideal no-attack model.

The resource estimates in Table~\ref{tab:qss_comparison} depend on the assumptions of each study. In particular, the reported efficiencies of Xing et al. and Tian et al. do not explicitly subtract all bits disclosed during reconstruction checks. Comparing qubit costs at a common security target would require consistent accounting of the verification procedures and their resource requirements.

\section{Conclusion}\label{sec:6}

In this work, we proposed a new three-party quantum secret sharing protocol based on a three-qubit W state for distributing a predetermined sequence of classical bits. Alice's Hadamard-gate choices and the measurement-basis choices of Bob and Charlie determine the information, test and semi-information rounds. In the ideal model without an attack, transmitting an $n$-bit secret requires an average of $3n$ protocol rounds, giving a qubit efficiency of $1/9$. We analyzed intercept-and-resend and entangle-and-measure attacks by an internal adversary targeting Charlie's qubit and derived the corresponding single-round and cumulative detection probabilities. For the entangle-and-measure attack, the analysis also relates the information available through the ancillary system to the probability of forbidden outcomes in the test rounds.

The protocol was implemented in Python using the Qiskit framework and evaluated through ideal simulation and experiments on IBM Quantum hardware. The simulation results are consistent with the theoretical predictions for round efficiency and attack detection, while the hardware measurements provide estimates of round efficiency, the bit error rate of the reconstructed secret, forbidden-outcome probabilities and the information obtained through the specified ancilla measurement. Since hardware noise can produce forbidden outcomes even without an attack, a calibrated statistical criterion was introduced to test for excess disturbance relative to the measured background. The detection and false-alarm curves for this criterion were calculated using a binomial model with experimentally estimated forbidden-outcome probabilities. The results show that the sensitivity of the criterion depends on the interaction parameter, the hardware background and the number of test rounds.

The distinctive contribution of the protocol lies in how secret transmission and tests for forbidden outcomes are combined within the same ensemble of encoded W states. The rounds used for testing are determined during transmission rather than designated in advance, while suitable semi-information rounds provide additional opportunities for secret transmission and improve efficiency. The comparison with related QSS and SQSS protocols places these features in the context of different receiver capabilities and verification procedures, while the reported qubit efficiencies must be interpreted under the resource-accounting assumptions of each scheme.

Together, these results connect the mathematical description of the protocol with its software implementation and a finite-sample statistical analysis based on quantum-hardware measurements. The security conclusions concern the analyzed attack models, while the hardware criterion relies on the stated statistical assumptions, including independent rounds and stationary device statistics. Further work could explore extensions of the protocol to larger numbers of participants and investigate a broader class of attacks by dishonest participants.

\par\medskip
\begingroup
\small
\noindent
\textbf{Acknowledgments.} The study was implemented in the framework of the Basic Research Program at the HSE University (HSE-BR-2025-007). 

\medskip

\noindent
\textbf{Data availability.} The source code, recorded IBM Quantum measurement data and generated figures are publicly available on GitHub \cite{QSSCode2026}.
\par
\endgroup

\begingroup
\setlength{\emergencystretch}{1em}
\setlength{\bibitemsep}{2pt}
\printbibliography

@article{Shamir1979,
  author  = {Shamir, Adi},
  title   = {How to Share a Secret},
  journal = {Communications of the ACM},
  volume  = {22},
  number  = {11},
  pages   = {612--613},
  year    = {1979},
  doi     = {10.1145/359168.359176}
}

@inproceedings{Shor1994,
  author    = {Shor, Peter W.},
  title     = {Algorithms for Quantum Computation: Discrete Logarithms and Factoring},
  booktitle = {Proceedings of the 35th Annual Symposium on Foundations of Computer Science},
  pages     = {124--134},
  year      = {1994},
  publisher = {IEEE},
  doi       = {10.1109/SFCS.1994.365700}
}

@inproceedings{BennettBrassard1984,
  author    = {Bennett, Charles H. and Brassard, Gilles},
  title     = {Quantum Cryptography: Public Key Distribution and Coin Tossing},
  booktitle = {Proceedings of the IEEE International Conference on Computers, Systems and Signal Processing},
  pages     = {175--179},
  year      = {1984},
  address   = {Bangalore, India}
}

@article{Ekert1991,
  author  = {Ekert, Artur K.},
  title   = {Quantum Cryptography Based on Bell's Theorem},
  journal = {Physical Review Letters},
  volume  = {67},
  number  = {6},
  pages   = {661--663},
  year    = {1991},
  doi     = {10.1103/PhysRevLett.67.661}
}

@article{DengLongLiu2003,
  author  = {Deng, Fu-Guo and Long, Gui-Lu and Liu, Xiao-Shu},
  title   = {Two-Step Quantum Direct Communication Protocol Using the Einstein-Podolsky-Rosen Pair Block},
  journal = {Physical Review A},
  volume  = {68},
  number  = {4},
  pages   = {042317},
  year    = {2003},
  doi     = {10.1103/PhysRevA.68.042317}
}

@article{Hillery1999,
  author  = {Hillery, Mark and Bu{\v{z}}ek, Vladim{\'i}r and Berthiaume, Andr{\'e}},
  title   = {Quantum Secret Sharing},
  journal = {Physical Review A},
  volume  = {59},
  number  = {3},
  pages   = {1829--1834},
  year    = {1999},
  doi     = {10.1103/PhysRevA.59.1829}
}

@article{Guo2003,
  author  = {Guo, Guo-Ping and Guo, Guang-Can},
  title   = {Quantum Secret Sharing without Entanglement},
  journal = {Physics Letters A},
  volume  = {310},
  number  = {4},
  pages   = {247--251},
  year    = {2003},
  doi     = {10.1016/S0375-9601(03)00074-4}
}

@article{Xiao2004,
  author  = {Xiao, Li and Long, Gui-Lu and Deng, Fu-Guo and Pan, Jian-Wei},
  title   = {Efficient Multiparty Quantum-Secret-Sharing Schemes},
  journal = {Physical Review A},
  volume  = {69},
  number  = {5},
  pages   = {052307},
  year    = {2004},
  doi     = {10.1103/PhysRevA.69.052307}
}

@inproceedings{Gao2005LocalOperations,
  author    = {Gao, Fei and Guo, Fen-Zhuo and Wen, Qiao-Yan and Zhu, Fu-Chen},
  title     = {Quantum Sharing of Classical Secret Based on Local Operations},
  booktitle = {Proceedings of the 5th International Conference on Information Communications and Signal Processing},
  pages     = {986--988},
  year      = {2005},
  address   = {Bangkok, Thailand},
  doi       = {10.1109/ICICS.2005.1689198}
}

@article{ZhangLiMan2005,
  author  = {Zhang, Zhan-Jun and Li, Yong and Man, Zhong-Xiao},
  title   = {Multiparty Quantum Secret Sharing},
  journal = {Physical Review A},
  volume  = {71},
  number  = {4},
  pages   = {044301},
  year    = {2005},
  doi     = {10.1103/PhysRevA.71.044301}
}

@article{ZhangMan2005,
  author  = {Zhang, Zhan-Jun and Man, Zhong-Xiao},
  title   = {Multiparty Quantum Secret Sharing of Classical Messages Based on Entanglement Swapping},
  journal = {Physical Review A},
  volume  = {72},
  number  = {2},
  pages   = {022303},
  year    = {2005},
  doi     = {10.1103/PhysRevA.72.022303}
}

@article{Karlsson1999,
  author  = {Karlsson, Anders and Koashi, Masato and Imoto, Nobuyuki},
  title   = {Quantum Entanglement for Secret Sharing and Secret Splitting},
  journal = {Physical Review A},
  volume  = {59},
  number  = {1},
  pages   = {162--168},
  year    = {1999},
  doi     = {10.1103/PhysRevA.59.162}
}

@article{DengTrojan2005,
  author  = {Deng, Fu-Guo and Li, Xi-Han and Zhou, Hong-Yu and Zhang, Zhan-Jun},
  title   = {Improving the Security of Multiparty Quantum Secret Sharing against Trojan Horse Attack},
  journal = {Physical Review A},
  volume  = {72},
  number  = {4},
  pages   = {044302},
  year    = {2005},
  doi     = {10.1103/PhysRevA.72.044302}
}

@article{Qin2007HBB,
  author  = {Qin, Su-Juan and Gao, Fei and Wen, Qiao-Yan and Zhu, Fu-Chen},
  title   = {Cryptanalysis of the Hillery-Buzek-Berthiaume Quantum Secret-Sharing Protocol},
  journal = {Physical Review A},
  volume  = {76},
  number  = {6},
  pages   = {062324},
  year    = {2007},
  doi     = {10.1103/PhysRevA.76.062324}
}

@article{Song2009ParticipantAttack,
  author  = {Song, Ting-Ting and Zhang, Jie and Gao, Fei and Wen, Qiao-Yan and Zhu, Fu-Chen},
  title   = {Participant Attack on Quantum Secret Sharing Based on Entanglement Swapping},
  journal = {Chinese Physics B},
  volume  = {18},
  number  = {4},
  pages   = {1333--1337},
  year    = {2009},
  doi     = {10.1088/1674-1056/18/4/007}
}

@article{DurVidalCirac2000,
  author  = {D{\"u}r, Wolfgang and Vidal, Guifr{\'e} and Cirac, J. Ignacio},
  title   = {Three Qubits Can Be Entangled in Two Inequivalent Ways},
  journal = {Physical Review A},
  volume  = {62},
  number  = {6},
  pages   = {062314},
  year    = {2000},
  doi     = {10.1103/PhysRevA.62.062314}
}

@article{Joo2005WState,
  author  = {Joo, Jaewoo and Park, Young Jai and Lee, Jinhyoung and Jang, Jingak and Kim, Inbo},
  title   = {Quantum Secure Communication via a W State},
  journal = {Journal of the Korean Physical Society},
  volume  = {46},
  number  = {4},
  pages   = {763--768},
  year    = {2005}
}

@article{XueYiCao2006,
  author  = {Xue, Zheng-Yuan and Yi, You-Min and Cao, Zhuo-Liang},
  title   = {Scheme for Sharing Classical Information via Tripartite Entangled States},
  journal = {Chinese Physics},
  volume  = {15},
  number  = {7},
  pages   = {1421--1424},
  year    = {2006},
  doi     = {10.1088/1009-1963/15/7/006}
}

@article{LiuTsaiHwang2012W,
  author  = {Liu, Lin-Lin and Tsai, Chia-Wei and Hwang, Tzonelih},
  title   = {Quantum Secret Sharing Using Symmetric W State},
  journal = {International Journal of Theoretical Physics},
  volume  = {51},
  number  = {7},
  pages   = {2291--2306},
  year    = {2012},
  doi     = {10.1007/s10773-012-1109-7}
}

@article{TsaiHwang2012,
  author  = {Tsai, Chia-Wei and Hwang, Tzonelih},
  title   = {Multi-party Quantum Secret Sharing Based on Two Special Entangled States},
  journal = {Science China Physics, Mechanics and Astronomy},
  volume  = {55},
  number  = {3},
  pages   = {460--464},
  year    = {2012},
  doi     = {10.1007/s11433-012-4633-9}
}

@article{ZhangLiuFangWang2007,
  author  = {Zhang, Zhan-Jun and Liu, Yi-Min and Fang, Ming and Wang, Dong},
  title   = {Multiparty Quantum Secret Sharing Scheme of Classical Messages by Swapping Qudit-State Entanglement},
  journal = {International Journal of Modern Physics C},
  volume  = {18},
  number  = {12},
  pages   = {1885--1901},
  year    = {2007},
  doi     = {10.1142/S0129183107011807}
}

@article{Boyer2007,
  author  = {Boyer, Michel and Kenigsberg, Dan and Mor, Tal},
  title   = {Quantum Key Distribution with Classical Bob},
  journal = {Physical Review Letters},
  volume  = {99},
  number  = {14},
  pages   = {140501},
  year    = {2007},
  doi     = {10.1103/PhysRevLett.99.140501}
}

@article{LiChanLong2010,
  author  = {Li, Qin and Chan, Wai-Hong and Long, Dong-Yang},
  title   = {Semiquantum Secret Sharing Using Entangled States},
  journal = {Physical Review A},
  volume  = {82},
  number  = {2},
  pages   = {022303},
  year    = {2010},
  doi     = {10.1103/PhysRevA.82.022303}
}

@article{LiQiuMateus2013,
  author  = {Li, Lvzhou and Qiu, Daowen and Mateus, Paulo},
  title   = {Quantum Secret Sharing with Classical Bobs},
  journal = {Journal of Physics A: Mathematical and Theoretical},
  volume  = {46},
  number  = {4},
  pages   = {045304},
  year    = {2013},
  doi     = {10.1088/1751-8113/46/4/045304}
}

@article{TsaiYangLee2019W,
  author  = {Tsai, Chia-Wei and Yang, Chun-Wei and Lee, Narn-Yih},
  title   = {Semi-quantum Secret Sharing Protocol Using W-state},
  journal = {Modern Physics Letters A},
  volume  = {34},
  number  = {27},
  pages   = {1950213},
  year    = {2019},
  doi     = {10.1142/S0217732319502134}
}

@article{LiAnonymousW2024,
  author  = {Li, Guo-Dong and Cheng, Wen-Chuan and Wang, Qing-Le and Cheng, Long and Mao, Ying and Jia, Heng-Yue},
  title   = {Enhanced Quantum Secret Sharing Protocol for Anonymous Secure Communication Utilizing W States},
  journal = {iScience},
  volume  = {27},
  number  = {6},
  pages   = {109836},
  year    = {2024},
  doi     = {10.1016/j.isci.2024.109836}
}

@article{Xing2025W,
  author  = {Xing, Kai and Lu, Rongbo and Liu, Sihai and Lan, Lu},
  title   = {An Efficient and Secure Semi-Quantum Secret Sharing Scheme Based on W State Sharing of Specific Bits},
  journal = {Entropy},
  volume  = {27},
  number  = {11},
  pages   = {1107},
  year    = {2025},
  doi     = {10.3390/e27111107}
}

@article{LinYangTsaiHwang2013,
  author  = {Lin, Jason and Yang, Chun-Wei and Tsai, Chia-Wei and Hwang, Tzonelih},
  title   = {Intercept-Resend Attacks on Semi-Quantum Secret Sharing and the Improvements},
  journal = {International Journal of Theoretical Physics},
  volume  = {52},
  pages   = {156--162},
  year    = {2013},
  doi     = {10.1007/s10773-012-1314-4}
}

@article{YangTsaiHwang2012,
  author  = {Yang, Chun-Wei and Tsai, Chia-Wei and Hwang, Tzonelih},
  title   = {Thwarting Intercept-and-Resend Attack on Zhang's Quantum Secret Sharing Using Collective Rotation Noises},
  journal = {Quantum Information Processing},
  volume  = {11},
  number  = {1},
  pages   = {113--122},
  year    = {2012},
  doi     = {10.1007/s11128-011-0236-z}
}

@article{QinTsoDai2019,
  author  = {Qin, Huawang and Tso, Raylin and Dai, Yuewei},
  title   = {Quantum Secret Sharing by Using Fourier Transform on Orbital Angular Momentum},
  journal = {IET Information Security},
  volume  = {13},
  number  = {2},
  pages   = {104--108},
  year    = {2019},
  doi     = {10.1049/iet-ifs.2018.5149}
}

@article{He2024,
  author  = {He, Fan and Xin, Xiangjun and Li, Chaoyang and Li, Fagen},
  title   = {Security Analysis of the Semi-Quantum Secret-Sharing Protocol of Specific Bits and Its Improvement},
  journal = {Quantum Information Processing},
  year    = {2024},
  volume  = {23},
  number  = {2},
  pages   = {51},
  doi     = {10.1007/s11128-023-04255-z}
}

@article{Chou2021,
  author  = {Chou, Yao-Hsin and Zeng, Guo-Jyun and Chen, Xing-Yu and Kuo, Shu-Yu},
  title   = {Multiparty Weighted Threshold Quantum Secret Sharing Based on the Chinese Remainder Theorem to Share Quantum Information},
  journal = {Scientific Reports},
  volume  = {11},
  pages   = {6093},
  year    = {2021},
  doi     = {10.1038/s41598-021-85703-7}
}

@article{Long2021,
  author  = {Long, Yinxiang and Zhang, Cai and Sun, Zhiwei},
  title   = {Standard (3, 5)-Threshold Quantum Secret Sharing by Maximally Entangled 6-Qubit States},
  journal = {Scientific Reports},
  volume  = {11},
  pages   = {22649},
  year    = {2021},
  doi     = {10.1038/s41598-021-01893-0}
}

@article{Chen2005,
  author  = {Chen, Yu-Ao and Zhang, An-Ning and Zhao, Zhi and Zhou, Xiao-Qi and Lu, Chao-Yang and Peng, Cheng-Zhi and Yang, Tao and Pan, Jian-Wei},
  title   = {Experimental Quantum Secret Sharing and Third-Man Quantum Cryptography},
  journal = {Physical Review Letters},
  volume  = {95},
  number  = {20},
  pages   = {200502},
  year    = {2005},
  doi     = {10.1103/PhysRevLett.95.200502}
}

@article{Schmid2005,
  author  = {Schmid, Christian and Trojek, Pavel and Bourennane, Mohamed and Kurtsiefer, Christian and Żukowski, Marek and Weinfurter, Harald},
  title   = {Experimental Single Qubit Quantum Secret Sharing},
  journal = {Physical Review Letters},
  volume  = {95},
  number  = {23},
  pages   = {230505},
  year    = {2005},
  doi     = {10.1103/PhysRevLett.95.230505}
}

@article{Gaertner2007,
  author  = {Gaertner, Sascha and Kurtsiefer, Christian and Bourennane, Mohamed and Weinfurter, Harald},
  title   = {Experimental Demonstration of Four-Party Quantum Secret Sharing},
  journal = {Physical Review Letters},
  volume  = {98},
  number  = {2},
  pages   = {020503},
  year    = {2007},
  doi     = {10.1103/PhysRevLett.98.020503}
}

@article{Bogdanski2008,
  author  = {Bogdanski, Jan and Rafiei, Nima and Bourennane, Mohamed},
  title   = {Experimental Quantum Secret Sharing Using Telecommunication Fiber},
  journal = {Physical Review A},
  volume  = {78},
  number  = {6},
  pages   = {062307},
  year    = {2008},
  doi     = {10.1103/PhysRevA.78.062307}
}

@article{Zhou2018,
  author  = {Zhou, Yaoyao and Yu, Juan and Yan, Zhihui and Jia, Xiaojun and Zhang, Jing and Xie, Changde and Peng, Kunchi},
  title   = {Quantum Secret Sharing Among Four Players Using Multipartite Bound Entanglement of an Optical Field},
  journal = {Physical Review Letters},
  volume  = {121},
  number  = {15},
  pages   = {150502},
  year    = {2018},
  doi     = {10.1103/PhysRevLett.121.150502}
}

@article{Williams2019,
  author  = {Williams, Brian P. and Lukens, Joseph M. and Peters, Nicholas A. and Qi, Bing and Grice, Warren P.},
  title   = {Quantum Secret Sharing with Polarization-Entangled Photon Pairs},
  journal = {Physical Review A},
  volume  = {99},
  number  = {6},
  pages   = {062311},
  year    = {2019},
  doi     = {10.1103/PhysRevA.99.062311}
}

@article{Liu2023,
  author  = {Liu, Shuaishuai and Lu, Zhenguo and Wang, Pu and Tian, Yan and Wang, Xuyang and Li, Yongmin},
  title   = {Experimental Demonstration of Multiparty Quantum Secret Sharing and Conference Key Agreement},
  journal = {npj Quantum Information},
  volume  = {9},
  pages   = {92},
  year    = {2023},
  doi     = {10.1038/s41534-023-00763-z}
}

@article{Qin2024,
  author  = {Qin, Yue and Cheng, Jialin and Ma, Jingxu and Zhao, Di and Yan, Zhihui and Jia, Xiaojun and Xie, Changde and Peng, Kunchi},
  title   = {Efficient and Secure Quantum Secret Sharing for Eight Users},
  journal = {Physical Review Research},
  volume  = {6},
  number  = {3},
  pages   = {033036},
  year    = {2024},
  doi     = {10.1103/PhysRevResearch.6.033036}
}

@article{Yan2026,
  author  = {Yan, Haoxiong and Zang, Allen and Grebel, Joel and Wu, Xuntao and Chou, Ming-Han and Andersson, Gustav and Conner, Christopher R. and Joshi, Yash J. and Li, Shiheng and Miller, Jacob M. and Povey, Rhys G. and Qiao, Hong and Chitambar, Eric and Cleland, Andrew N.},
  title   = {Quantum Secret Sharing in a Triangular Superconducting Quantum Network},
  journal = {npj Quantum Information},
  year    = {2026},
  doi     = {10.1038/s41534-026-01341-9}
}

@misc{QISkit,
  author        = {Javadi-Abhari, Ali and
                   Treinish, Matthew and
                   Krsulich, Kevin and
                   Wood, Christopher J. and
                   Lishman, Jake and
                   Gacon, Julien and
                   Martiel, Simon and
                   Nation, Paul D. and
                   Bishop, Lev S. and
                   Cross, Andrew W. and
                   Johnson, Blake R. and
                   Gambetta, Jay M.},
  title         = {Quantum computing with {Q}iskit},
  year          = {2024},
  doi           = {10.48550/arXiv.2405.08810},
  eprint        = {2405.08810},
  archivePrefix = {arXiv},
  primaryClass  = {quant-ph}
}

@article{Diogo2019,
  author  = {Cruz, Diogo and Fournier, Romain and Gremion, Fabien and Jeannerot, Alix and Komagata, Kenichi and Tosic, Tara and Thiesbrummel, Jarla and Chan, Chun Lam and Macris, Nicolas and Dupertuis, Marc-André and Javerzac-Galy, Clément},
  title   = {Efficient Quantum Algorithms for GHZ and W States, and Implementation on the IBM Quantum Computer},
  journal = {Advanced Quantum Technologies},
  volume  = {2},
  number  = {5-6},
  pages   = {1900015},
  doi     = {10.1002/qute.201900015},
  year    = {2019}
}

@article{Xie2015Specific,
  author  = {Xie, Chen and Li, Lvzhou and Qiu, Daowen},
  title   = {A Novel Semi-Quantum Secret Sharing Scheme of Specific Bits},
  journal = {International Journal of Theoretical Physics},
  volume  = {54},
  pages   = {3819--3824},
  year    = {2015},
  doi     = {10.1007/s10773-015-2622-2}
}

@article{Tian2021Specific,
  author  = {Tian, Yuan and Li, Jian and Chen, Xiu-Bo and Ye, Chong-Qiang and Li, Heng-Ji},
  title   = {An Efficient Semi-Quantum Secret Sharing Protocol of Specific Bits},
  journal = {Quantum Information Processing},
  volume  = {20},
  pages   = {217},
  year    = {2021},
  doi     = {10.1007/s11128-021-03157-2}
}

@article{Li2021Cluster,
  author  = {Li, Chaoyang and Ye, Chongqiang and Tian, Yuan and Chen, Xiu-Bo and Li, Jian},
  title   = {Cluster-State-Based Quantum Secret Sharing for Users with Different Abilities},
  journal = {Quantum Information Processing},
  volume  = {20},
  pages   = {385},
  year    = {2021},
  doi     = {10.1007/s11128-021-03327-2}
}

@article{Chen2022Chi,
  author  = {Chen, Ying and Ye, Tian-Yu},
  title   = {Semiquantum Secret Sharing by Using {$\chi$}-Type States},
  journal = {The European Physical Journal Plus},
  volume  = {137},
  pages   = {1331},
  year    = {2022},
  doi     = {10.1140/epjp/s13360-022-03521-w}
}

@article{Hou2024Circular,
  author  = {Hou, Yan-Yan and Xu, Tao and Li, Jian and Ye, Chong-Qiang and Wang, Z. and Liu, Xin-Yu},
  title   = {Circular Semi-Quantum Secret Sharing Based on Hybrid Single Particle and {GHZ}-Type States},
  journal = {Laser Physics Letters},
  volume  = {21},
  pages   = {025202},
  year    = {2024},
  doi     = {10.1088/1612-202X/ad1aab}
}

@article{Hoeffding1963,
  author  = {Hoeffding, Wassily},
  title   = {Probability Inequalities for Sums of Bounded Random Variables},
  journal = {Journal of the American Statistical Association},
  year    = {1963},
  volume  = {58},
  number  = {301},
  pages   = {13--30},
  doi     = {10.1080/01621459.1963.10500830}
}

@article{YinFu2016Xie,
  author  = {Yin, Aihan and Fu, Fangbo},
  title   = {Eavesdropping on Semi-Quantum Secret Sharing Scheme of Specific Bits},
  journal = {International Journal of Theoretical Physics},
  year    = {2016},
  volume  = {55},
  number  = {9},
  pages   = {4027--4035},
  doi     = {10.1007/s10773-016-3031-x}
}

@article{Ban1997,
  author  = {Ban, Masashi and Yamazaki, Kouichi and Hirota, Osamu},
  title   = {Accessible information in combined and sequential quantum measurements on a binary-state signal},
  journal = {Physical Review A},
  volume  = {55},
  number  = {1},
  pages   = {22--26},
  year    = {1997},
  doi     = {10.1103/PhysRevA.55.22}
}

@article{KaurKumar2018,
  author  = {Kaur, Hargeet and Kumar, Atul},
  title   = {Analyzing the efficiency of partially entangled states in {Vaidman}'s-type games and its applications in {Quantum Secret Sharing}},
  journal = {International Journal of Computers and Applications},
  year    = {2019},
  volume  = {41},
  number  = {1},
  pages   = {2--13},
  doi     = {10.1080/1206212X.2018.1465666}
}

@misc{QSSCode2026,
  author       = {Teregulov, Timur},
  title        = {Three-party {W}-state quantum secret sharing: source code and measurement data},
  year         = {2026},
  howpublished = {GitHub},
  url          = {https://github.com/iskrok/quantum_source_code/tree/0e1fa1b68828f234bcf728ddbf32e5268434b30b/w_state_qss}
}
\endgroup

\end{document}